\documentclass[journal]{IEEEtran}
\usepackage{amsfonts}
\usepackage{verbatim}
\usepackage{mathrsfs}
\usepackage{amssymb,amsmath}
\usepackage[noblocks]{authblk}
\usepackage{stfloats}
\usepackage{algorithm}
\usepackage{algorithmic}
\usepackage{cite,subfigure,graphicx,wrapfig,verbatim,amsmath,amsfonts,amssymb,color, epsfig,graphics}
\DeclareGraphicsExtensions{.pdf,.jpeg,.png,.jpg}
\usepackage{theorem}
\usepackage{array,color}
\usepackage{multirow}
\usepackage{makecell}
\usepackage[hyphens]{url}
\usepackage{breakurl}

\theoremheaderfont{\normalfont\bfseries}

\newcommand{\eg}{e.g.}

\begin{document}
\title{Vehicle as a Service (VaaS): Leverage Vehicles to Build Service Networks and Capabilities for Smart Cities
}

\author{\IEEEauthorblockN{Xianhao Chen,~\IEEEmembership{Member,~IEEE}, Yiqin Deng,~\IEEEmembership{Member,~IEEE}, Haichuan Ding,~\IEEEmembership{Member,~IEEE}, Guanqiao Qu, ~\IEEEmembership{Graduate Student Member,~IEEE}, Haixia Zhang,~\IEEEmembership{Senior Member,~IEEE}, Pan Li,~\IEEEmembership{Senior Member,~IEEE}, Yuguang Fang,~\IEEEmembership{Fellow,~IEEE}
}

\thanks{Xianhao Chen and Guanqiao Qu are with the Department of Electrical and Electronic Engineering, University of Hong Kong, Pok Fu Lam, Hong Kong, China (e-mail: xchen@eee.hku.hk; gqqu@eee.hku.hk).}
\thanks{Yiqin Deng and Haixia Zhang are with the School of Control Science and Engineering and with the Shandong Key Laboratory of Wireless Communication Technologies, Shandong University, Jinan 250061, Shandong, China (e-mail: yiqin.deng@email.sdu.edu.cn; haixia.zhang@sdu.edu.cn).}
\thanks{Haichuan Ding is with the School of Cyberspace Science and Technology, Beijing Institute of Technology, Beijing 100081, China
(email: hcding@bit.edu.cn).}
\thanks{Pan Li is with the Department of Electrical Engineering and Computer Science, Case Western Reserve University, Cleveland, OH 44106, USA (email: lipan@case.edu).}
\thanks{Yuguang Fang is with the Department of Computer Science, City University of Hong Kong, Kowloon, Hong Kong, China (e-mail: my.fang@cityu.edu.hk).}

}


\maketitle

\begin{abstract}
Smart cities demand resources for rich immersive \underline{s}ensing, ubiquitous \underline{c}ommunications, powerful \underline{c}omputing, large \underline{s}torage, and high \underline{i}ntelligence (SCCSI) to support various kinds of applications, such as public safety, connected and autonomous driving, smart and connected health, and smart living. At the same time, it is widely recognized that vehicles such as autonomous cars, equipped with significantly powerful SCCSI capabilities, will become ubiquitous in future smart cities. By observing the convergence of these two trends, this article advocates the use of vehicles to build a cost-effective service network, called the Vehicle as a Service (VaaS) paradigm, where vehicles empowered with SCCSI capability form a web of mobile servers and communicators to provide SCCSI services in smart cities. Towards this direction, we first examine the potential use cases in smart cities and possible upgrades required for the transition from traditional vehicular ad hoc networks (VANETs) to VaaS. Then, we will introduce the system architecture of the VaaS paradigm and discuss how it can provide SCCSI services in future smart cities, respectively. At last, we identify the open problems of this paradigm and future research directions, including architectural design, service provisioning, incentive design, and security \& privacy. We expect that this paper paves the way towards developing a cost-effective and sustainable approach for building smart cities. 

\end{abstract}


\section{Introduction}
Imagine Alice visits a completely strange tourist city full of attractions. It would be nice if she could automatically connect to a network from which she could get any personalized information and recommendations at her fingertips. For example, she may simply ask her mobile device ``please recommend me a restaurant with waiting time shorter than 5 minutes.'' The mobile device, which is unsurprisingly ``smart'' enough to have her food preference (\textit{personalized recommendation systems}) and her precise location information (\textit{localization}), searches for a suitable restaurant by comparing real-time information, such as restaurants' recipes and the estimated waiting time with remarkable precision. After making the decision, the mobile device automatically contacts the nearest shared self-driving car, which will drive her to the destination safely, timely, and smoothly (\textit{connected and automated driving}). Throughout this journey, Alice can use her mobile device or the car's cameras to access information about her surroundings in real-time, thanks to \textit{augmented reality}. Finally, the car would park at the closest available parking spot near the restaurant, thanks to \textit{parking availability detection}.

The above story is just a glimpse of use cases of smart cities~\cite{hong2020hong,zanella2014internet}. To many, these scenarios also fall into the category called Internet of Things (IoT). According to Hong Kong Smart City Blueprint 2.0~\cite{hong2020hong}, a smart city intends to make the city provide the following ``smart'' features: smart mobility, smart living, smart environment, smart people, smart government, and smart economy, which intend to improve people’s quality of life (QoL). The ultimate goal is to provide its residents and visitors with more convenient daily routines, better lifestyles and living, improved healthcare and aging, and better inhabiting environments, which cannot be achieved by a single sector or a single city authority, but by concerted and collective efforts from all participants. Thus, smart cities demand resources for rich immersive sensing, ubiquitous communications, powerful computing, large storage, and high intelligence (SCCSI) to support various kinds of applications.

To fulfill the envisioned applications, a diverse set of sensing devices should be deployed sufficiently many premises to sense the ``pulse of the city''. The generated big data has to be transported to certain locations where they will be consumed, temporarily stored, or further processed for intelligence extraction in a timely fashion. Without doubt, the vision of a smart city cannot be achieved without a coordinated and concerted supporting service network for provisioning SCCSI resources and derived services. In this regard, 5G and beyond (5G+) is widely regarded as the mainstream solution~\cite{letaief2019roadmap,Samsung6G}. Unlike previous generations of wireless cellular systems that are mainly for data delivery, 5G+ systems feature dense deployments of (small) base stations and localized edge clouds to provide more effective communications, computing, and data storage services in close proximity of interconnected IoT devices, thereby forming a web of SCCSI capabilities to not only connect but also compute in a city.

Although 5G+ has a highly attractive blueprint, regrettably, the high Capital Expenditure (CAPEX) and Operational Expenditure (OPEX) have become bottlenecks~\cite{habibi2019comprehensive}. For example, to shift computation from end devices to the network, heterogeneous base stations, ranging from macro base stations (BSs) to small base stations (SBS), are envisioned to be equipped with edge computing nodes to provide low-latency and high-throughput edge learning/inference services, which heavily reckons on expensive GPU servers. Such infrastructure constructions (cellular backbones, BSs, SBSs, and edge nodes) will be costly, which is particularly true when considering that all these network entities have to be conformal to the operational standards for energy hungry 5G+~\cite{shurdi20215g,zhang2016fundamental,wu2017overview}. Moreover, if a city elects a cellular operator to build cellular infrastructure and run the smart city operations, OPEX for a day-to-day operational expense, such as the cost of workers, facility upgrade (e.g., GPU server upgrade), rent, and electricity, is also known to be highly expensive in 5G+~\cite{lopez2022survey}. Thus, a critical question is: is there any alternative economically sound approach? 

The answer to this question is positive and this position paper aims to provide such an alternative or a complementary solution to 5G+, which is called ``vehicle as a service (VaaS)''. It is commonly observed that the most visible mobile things in most cities are probably vehicles, on the ground (\eg, cars, trucks, public transits) and in the air (\eg, airships, balloons, unmanned aerial vehicles (UAVs)). In the future, connected and autonomous vehicles (CAVs) will fundamentally revolutionize the way of transport in our daily lives. However, we anticipate that the impact of CAVs can go far beyond transportation. On the one hand, \textit{vehicles will be equipped with significantly powerful SCCSI resources to enable self-driving and other in-car infotainments, and these resources will be severely underutilized most of the time, creating a staggering amount of spare resources}. For example, cars are parked for 23 hours per day on average ~\cite{RAC2021cars}. Even for vehicles on the move, say, when they are caught in a traffic jam or sparsely distributed traffic flow, there are severely underutilized resources given the varying environments and heterogeneous onboard computing capabilities. Turning millions of powerful autonomous vehicles and/or vehicles empowered with powerful SCCSI capabilities into mobile servers and communicators will fundamentally revolutionize the telecommunication and computing industries without adding significant investment cost in infrastructure and subsequent operational and maintenance cost, which could potentially offer a disruptive technologies in smart cities. On the other hand, \textit{the omnipresent vehicles can easily get really close to any things (people or machines) for proximity service provisioning, which is exactly aligned well to the 5G+ trend that both communications and computing capabilities are being pushed to the network edge to serve end users as locally as possible}~\cite{garcia2021tutorial,you2021towards,wang2023road}. The service provisioning on premises can not only reduce service delay (latency) but also relieves the communication burden of base stations and hence backbone network as a large portion of data would not need to be uploaded to base stations or roadside edge nodes for computing.



With the pool of significantly powerful SCCSI resources equipped on vehicles, VaaS can beef up smart cities/IoT applications in almost every aspect. Such vehicles can provide sensing services by sharing their onboard sensory information with other interested parties, relay/transport data for other vehicles/citizens, act as edge servers to process the massive data from the pervasive roadside IoT devices, and store the data for fast data retrieval for other vehicles/citizens. Importantly, the SCCSI service network is ``service-oriented'' and features the convergence of SCCSI as many emerging services involve multi-type multi-dimensional resources. Using the example at the beginning, calculating the travel and waiting time for Alice requires an integrated design of sensing, communications, and intelligence. Clearly, achieving the design goal requires a holistic design approach: (1) network-wide SCCSI resources on vehicles and on existing infrastructure must be orchestrated in an interactive, collaborative, and concerted manner with the ultimate goal of elevating users' service experience; (2) Incentives must be created for vehicles to be compensated for the costs of their resources; (3) Security and privacy issues must be carefully addressed in order to deliver trustworthy services. While conceptual studies for vehicular networks have emerged in recent years in one way or the other, the research on VaaS in line of our disruptive thinking is still in its fancy. This paper serves as a starting point discussing critical design issues for VaaS, an attempt to envision its global picture, a survey of the relevant works, and more importantly, a call for the creation of such an architecture.

As a final remark, VaaS has not ruled out the interactions with SCCSI-empowered fixed edge infrastructure, such as 5G+ infrastructure. In fact, VaaS will proactively place such a partial infrastructure, such as BSs and SBSs, access points (APs) in WiFi, and roadside units (RSUs) in intelligent transportation systems (ITS), with SCCSI-empowered capability to form a globally effective network for SCCSI services. \textit{The resulting network, with both mobile and fixed infrastructure, is termed as the ``SCCSI service network'' in this paper}\footnote{In this paper, we will use VaaS as the concept of leveraging vehicles to build up service networks and capabilities for smart cities, while using ``SCCSI service network'' to refer to the proposed network architecture for SCCSI resources used to provide services and capabilities to support smart cities in Section \ref{architecture}. However, we will also use them interchangeably whenever there is no confusion.}. Unlike our proposal, cellular operators merely emphasize the use of 5G+ infrastructure for digital intelligentization. For example, China Mogo launched one of the biggest autonomous driving projects and recently demonstrated that autonomous driving can operate normally by relying on roadside 5G infrastructure, even if all intelligence features in vehicles are turned off~\cite{avery2021china}. If we treat traditional autonomous driving targets leveraging in-vehicle capabilities while Mogo and alike emphasize the use of fixed infrastructure, the SCCSI service network strikes in between and attempts to take advantage of both fixed and mobile infrastructures to boost the performance of autonomous driving, intelligent traffic control, and other important value-added services and use cases (e.g., public safety, smart health, and smart living) in smart cities.

\subsection{Our Contributions and Comparisons with Prior Surveys}
While vehicular edge computing has received some attention in recent years, the majority of works focus on employing edge servers, such as 5G+ base stations or roadside units co-located with computing servers, to provide edge services to vehicles. In comparison, the leverage of vehicles as mobile edge infrastructure, which is the focus of this paper, has received less attention. To start with, this position paper takes a top-down approach to review the key design issues of VaaS. We first advocate a supporting service framework and introduce the critical components. Then, we present service provisioning from five categories, i.e., sensing, communications, computing, storage, and intelligence, respectively, in VaaS. In each category, we review the relevant literature and identify the critical research problems. 

\begin{table*}[htbp]\label{table1}
\centering
\caption{Summary of the related surveys/articles.}
  \renewcommand{\arraystretch}{1.4}{
  \setlength{\tabcolsep}{2mm}{
\begin{tabular}{c|c|c|c|c|c|c|c}
\hline
\textbf{References}                                         & \textbf{Year}&\begin{tabular}[c]{@{}c@{}}\textbf{Description}\end{tabular} & \begin{tabular}[c]{@{}c@{}}\textbf{Focus on}\\\textbf{vehicles as}\\\textbf{resources}\end{tabular}
& \begin{tabular}[c]{@{}c@{}}\textbf{Literature}\\\textbf{review}\end{tabular} & \begin{tabular}[c]{@{}c@{}}\textbf{Discussions}\\\textbf{on SCCSI}\\\textbf{convergence}\end{tabular} & \begin{tabular}[c]{@{}c@{}}\textbf{Architectural}\\\textbf{design}\end{tabular}  & \begin{tabular}[c]{@{}c@{}}\textbf{Network}\\\textbf{edge}\end{tabular}\\ \hline
\begin{tabular}[c]{@{}c@{}}\cite{raza2019survey}\end{tabular}    & 2019     &  \begin{tabular}[c]{@{}c@{}}Illustrates the architecture of vehicular edge computing,\\ along with its services and relevant technical solutions.\end{tabular}& $\times$  & $\checkmark$  & $\times$   & $\times$  & $\checkmark$    \\ \hline
\cite{zhang2019mobile}  & 2019 & 
 \begin{tabular}[c]{@{}c@{}}
Surveys the latest development
in edge systems for\\Internet of Vehicles, including edge computing,\\ edge caching, and edge AI.\end{tabular}
& $\times$ & $\checkmark$ & $\times$ & $\times$ & $\checkmark$ \\ \hline
\cite{dziyauddin2019computation} & 2019 & \begin{tabular}[c]{@{}c@{}}
Reviews the computation offloading and caching\\ problems in vehicular edge computing systems.\end{tabular}
& $\times$ & $\checkmark$ & $\times$ & $\times$ & $\checkmark$ \\ \hline
\cite{alhilal2020distributed} & 2020  & \begin{tabular}[c]{@{}c@{}}
Provides an overview of edge computing for vehicles\\ and conducts real-world case studies.\end{tabular}& $\times$  & $\checkmark$  & $\times$  & $\times$ & $\checkmark$\\ \hline
\cite{abuelela2010taking} & 2010  & \begin{tabular}[c]{@{}c@{}} Proposes the concept of vehicular cloud computing,\\ where underutilized vehicles' resources can be\\ shared between drivers or rented over the Internet to\\ various customers.\end{tabular}
& $\checkmark$  & $\times$  & $\times$  & $\times$ & $\times$\\ \hline
\cite{eltoweissy2010towards} & 2010  & \begin{tabular}[c]{@{}c@{}}Coins the term, Autonomous Vehicular Clouds (AVCs),\\ where mobile AVC resources can be pooled\\ dynamically to serve authorized users.\end{tabular}
& $\checkmark$  & $\times$  & $\times$  & $\times$ & $\times$\\ \hline
\cite{whaiduzzaman2014survey}  & 2014 & \begin{tabular}[c]{@{}c@{}} Presents a taxonomy for vehicular cloud computing \\ with an emphasis on the applications, cloud formations,\\ key management, inter-cloud communication\\ systems, and privacy and security issues.\end{tabular}
& $\checkmark$  & $\checkmark$  & $\times$  & $\times$ & $\times$\\ \hline
\cite{abdelhamid2015vehicle}   & 2015
&  \begin{tabular}[c]{@{}c@{}}
Introduce the concept of Vehicle as a Resource (VaaR)\\ and elaborate on how vehicles' resources can be\\ used for service provisioning.\end{tabular}
& $\checkmark$  & $\times$  & $\times$  & $\times$ & $\checkmark$\\ \hline
\cite{altintas2015making} 
& 2015 & \begin{tabular}[c]{@{}c@{}}
Advocates the use of cars as ICT infrastructure for\\ service provisioning.\end{tabular}  & $\checkmark$  & $\times$  & $\times$  & $\times$ & $\checkmark$\\ \hline
\cite{hou2016vehicular}  & 2016 & \begin{tabular}[c]{@{}c@{}}Conceives the idea of utilizing vehicles as infrastructure\\ for communication and computing service provisioning,\\ and carries out quantitative analysis of the framework.\end{tabular}
& $\checkmark$  & $\times$  & $\times$  & $\times$ & $\checkmark$\\ \hline
\cite{boukerche2018vehicular}  & 2018 & \begin{tabular}[c]{@{}c@{}} Reviews the approaches and solutions for \\vehicular cloud computing, featuring applications,\\ services, and traffic models that enable\\ vehicular clouds in a dynamic environment.
\end{tabular}
& $\checkmark$  & $\checkmark$  & $\times$  & $\times$ & $\times$\\ \hline
Ours  & 2023 & \begin{tabular}[c]{@{}c@{}} Presents a concept ``vehicle as a service'', presents\\ the architectural design of the SCCSI service networks, \\elaborates on numerous use cases in smart cities,\\ and reviews the related literature in SCCSI aspects.\end{tabular}
& $\checkmark$  & $\checkmark$  & $\checkmark$  & $\checkmark$ & $\checkmark$\\ \hline
\end{tabular}}}
\end{table*}


\begin{figure*}
\centering
\includegraphics[width=0.6\textwidth]{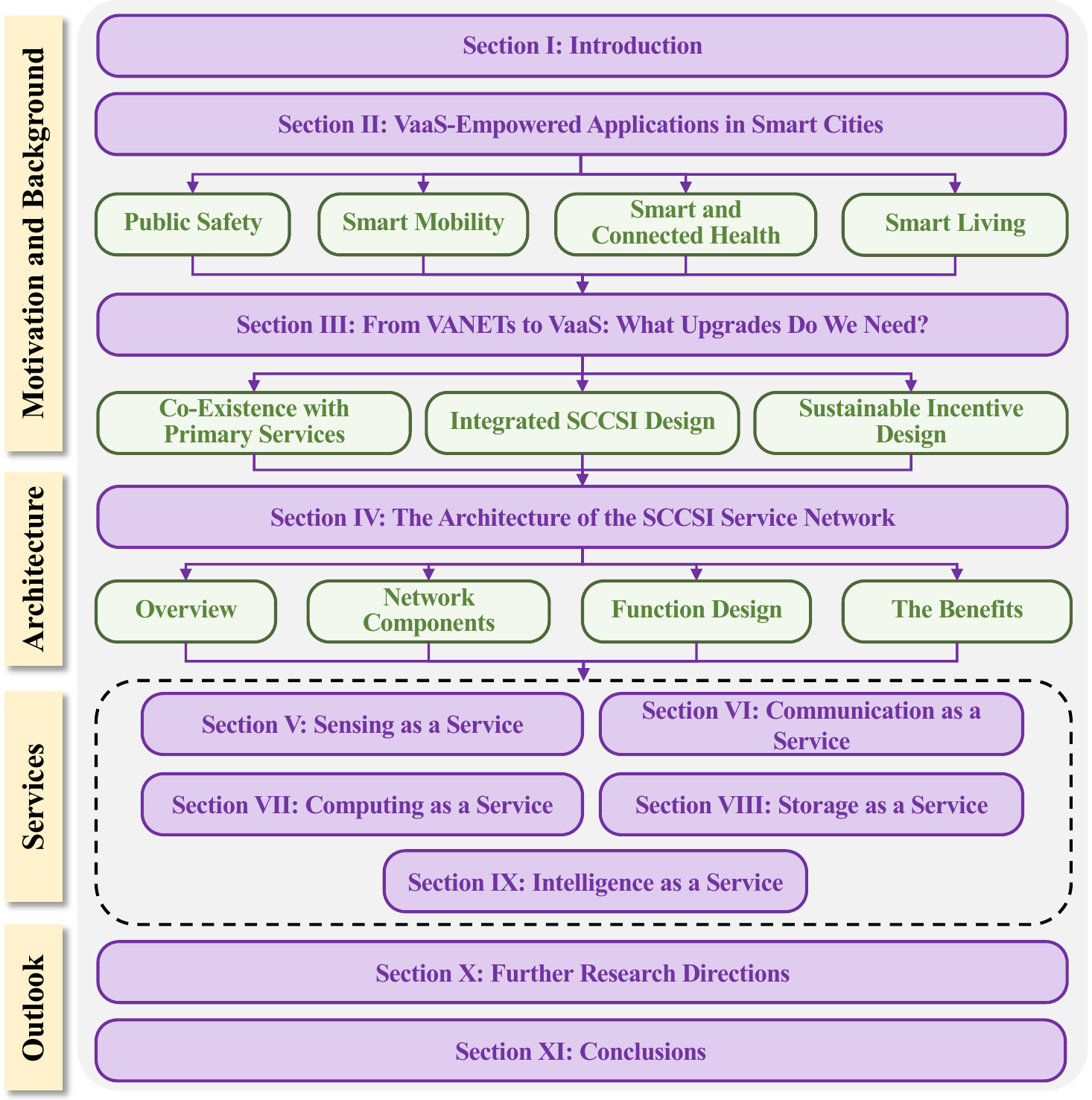}
    \caption{The organization of the paper and the interdependencies between different sections.}\label{structure}
\vspace*{-0.1in}
\end{figure*}

This paper is considerably different from prior surveys on vehicular networking, such as \cite{karagiannis2011vehicular,chen2022vehicular}, because the latter only addresses networking issues. This paper is also different from existing surveys on vehicular edge computing~\cite{raza2019survey,zhang2019mobile,dziyauddin2019computation} due to different focuses. Although some of these articles, such as \cite{abdelhamid2015vehicle,zhang2019mobile,dziyauddin2019computation,alhilal2020distributed}, also discuss the leverage of vehicles to serve as edge servers, there is no survey dedicated to offering a systematic design of VaaS. This survey also differs from the existing articles/surveys on vehicular cloud computing\cite{liao2019fog,altintas2015making,hou2016vehicular,abuelela2010taking,eltoweissy2010towards,olariu2011taking}, which utilize vehicles' resources as clouds to support computing over the Internet. The most related works might be 
\cite{altintas2015making,abdelhamid2015vehicle,hou2016vehicular}, which advocates the leverage of vehicles as ICT resources at the network edge. Nevertheless, these works have not conducted a literature review on the related topics and lack a proposal on the architectural design. To be specific, we identify the upgrades needed for the transition from VANETs with single-dimensional resource features to VaaS with multi-dimensional resource features, propose the promising architectural design of VaaS to empower multi-dimensional SCCSI service provisioning, and conduct a literature review comprehensively covering the aspects of SCCSI. To our best knowledge, this paper is \textit{the first survey to comprehensively review and discuss the use of vehicles as mobile edge servers and communicators to form an SCCSI service network, in which multi-dimensional resources and services are jointly considered.} The comparisons with other related surveys/articles are outlined in Table \ref{table1}. The main contributions of this paper are summarized as follows.
\begin{itemize}
    \item We coin the concept of ``vehicle as a service'' where vehicles form a SCCSI service network for service provisioning in smart cities. Note that the vehicles act as multi-dimensional resource suppliers rather than end users as usually considered  in the current literature. 
    \item We identify the upgrades needed for traditional VANETs and propose the supporting architecture for VaaS.
    \item We conduct a literature survey on the aspects of sensing, communications, computing, storage, and intelligence, respectively. To our best knowledge, this is the first survey that comprehensively reviews and discusses the use of vehicles as mobile edge servers and communicators to form a service network to provide various services.
\end{itemize}

The remainder of this paper is organized as follows. We will first identify possible upgrades we need from VANETs to VaaS in Section \ref{limitations}. In Section \ref{architecture}, we propose the architecture and basic components of VaaS. With the supporting architecture in mind, in Section \ref{UseCases}, we present some relevant smart city applications powered by VaaS. From Section \ref{sensing} to \ref{intelligence}, we cover the aspects of sensing, communications, computing, storage, and intelligence in VaaS, with an emphasis on their tight coupling. In Section \ref{future}, we discuss the open problems, challenges, and research opportunities. In Section \ref{conclusion}, we conclude this paper. The structure of the paper is also outlined in Fig. \ref{structure}.

\section{VaaS-Empowered Applications in Smart Cities\label{UseCases}}
\begin{figure*}
\centering
\includegraphics[width=1\textwidth]{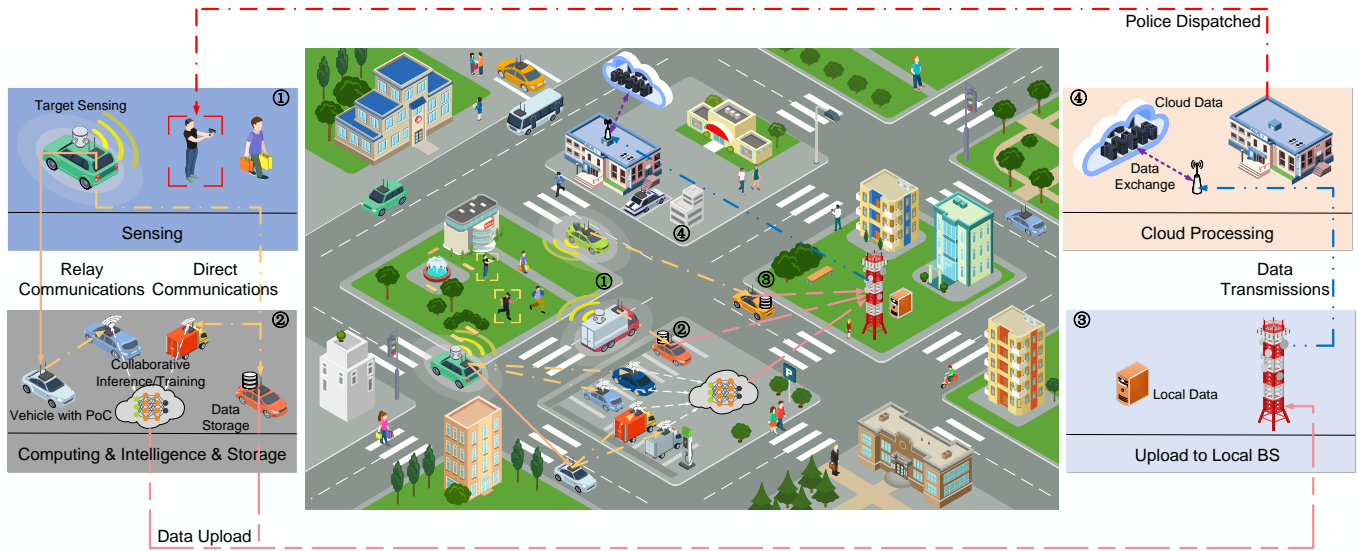}
    \caption{VaaS for public safety. In the figure, a crime scene is video-recorded from different angles and collaboratively analyzed by surrounding vehicles (both moving and parked vehicles), based on multi-modal sensing and collaborative perception. The summarized results will be transmitted to the police department for further action.   
    }\label{fig:publicsafety}
\vspace*{-0.1in}
\end{figure*}



SCCSI-empowered vehicles, acting as the mobile edge, can power a variety of applications. In what follows, we will elaborate on several use cases of our envisioned VaaS-enabled smart city.


\subsection{Public Safety} 
Public safety is always one of the primary concerns for residents and visitors in a smart city~\cite{pang2019spath}. VaaS can improve public safety by serving as a city-wide safety net, a surveillance system contributed by its citizens, as plotted in Fig. \ref{fig:publicsafety}. Each participating SCCSI-empowered vehicle in the network can serve as a sensing, communications, and computing device. For example, a crime scene can be video-recorded without the awareness of criminals. The video contents captured from different angles can be collaboratively analyzed by surrounding vehicles (either moving or parked vehicles) based on collaborative perception\cite{liu2020when2com}, after which the vehicles will trigger warnings to the criminals to deter their criminal acts while at the same time uploading the summarized results to the police department to take timely actions. One additional benefit is that, instead of dispatching police officers to the crime scene, SCCSI-empowered vehicles in the proximity can be alerted and incentivized to move to the proximity of the crime scene to help deter the crime progression (e.g., by triggering warnings), which could save police work load for other more important tasks.  

The envisioned VaaS can also prevent catastrophic incidents, such as crowd crush, in festival gatherings, which is particularly important for metropolitan cities with hot spots. The crowd collapse tragedy that occurred in Seoul during the 2022 Halloween festival killed at least 159 people. Unfortunately, this is not an isolated case. Similar casualties include Shanghai stampede in 2014 and Hong Kong stampede in 1993. To prevent the deadly incidents from occurrence, SCCSI-empowered vehicles can monitor crowd movement and conduct real-time simulations to forecast the possible outcomes. Once potential risks are detected, warnings and evacuation alerts will be sent to both the crowd and police stations to prevent disastrous outcomes.

\subsection{Smart Mobility}
It is unsurprising that VaaS helps smart mobility. Since this survey focuses on the aspect that vehicles serve as networked ``helpers'', we concentrate more on how vehicles can contribute their resources and collect data to help each other and the city to establish an ITS.

\begin{figure}
\centering
\includegraphics[width=0.45\textwidth]{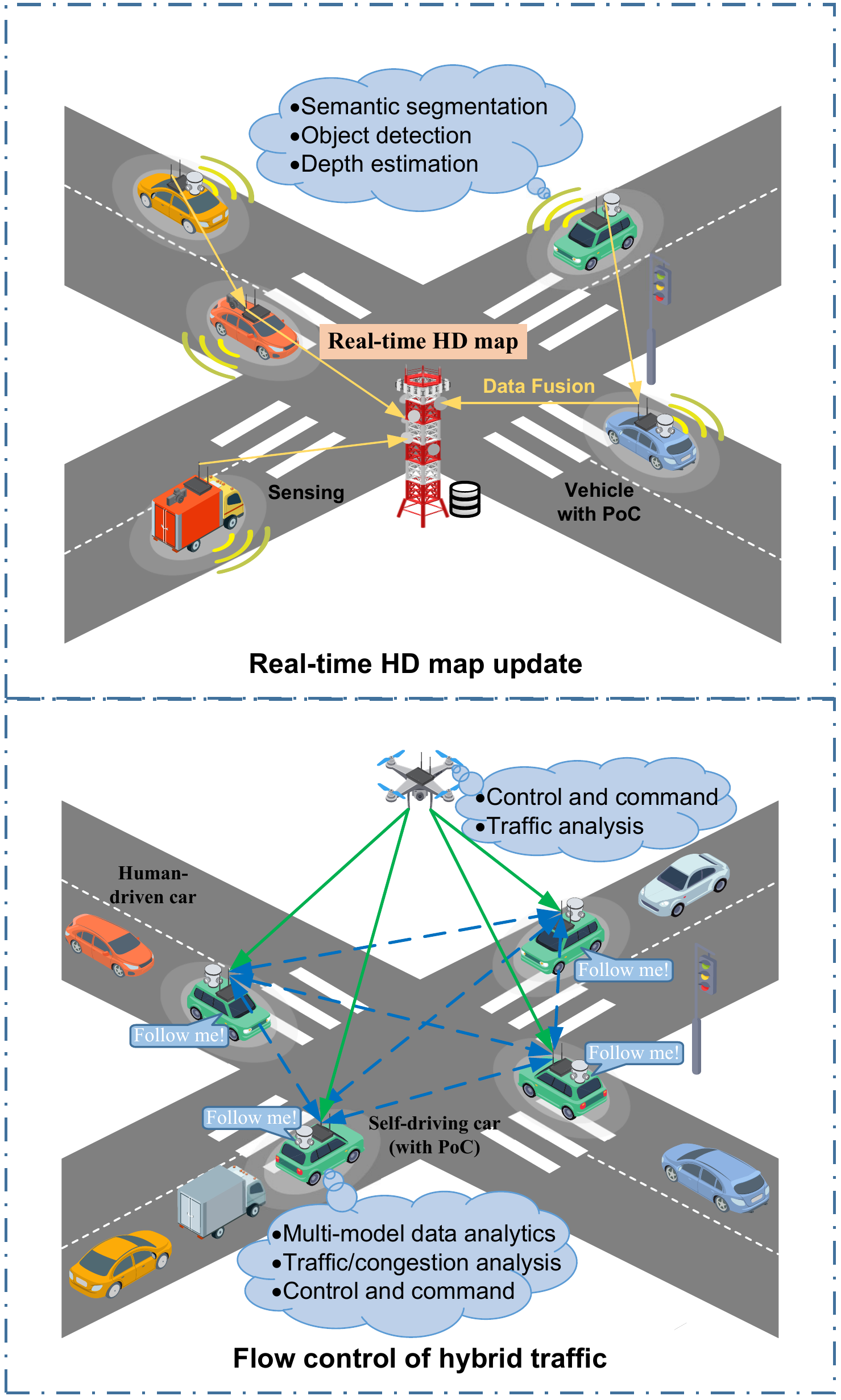}
    \caption{VaaS for smart mobility. In the upper figure, SCCSI-empowered vehicles analyze their sensory data (e.g., point cloud) and upload the summarized results to a base station, which fuses the data for real-time HD map update. In the bottom figure, SCCSI-empowered vehicles proactively regulate the traffic to relieve potential congestion.
    }\label{fig:smartmobility}
\vspace*{-0.1in}
\end{figure}

Enhancing traffic safety and efficiency can not only save lives and improve the quality of life but also decreases energy consumption and carbon dioxide emissions. To achieve this, vehicles must make informed decisions based on timely data all around. However, it is hard to gather data in a timely fashion due to the lack of sufficient sensing, communications, and computing capabilities in the current existing telecommunication infrastructure and/or ITS. Yet, with the envisioned VaaS, fine-grained traffic information can be easily acquired from SCCSI-empowered vehicles, which can then be forwarded to edge servers (either base stations, RSUs or parked roadside vehicles) for data aggregation and analytics. For instance, based on the information collected from surrounding vehicles, an edge server can identify and locate road hazards (e.g., broken glasses, rocks, falling trees), thereby updating a real-time high-definition (HD) map to inform incoming vehicles, which is crucial to road safety and traffic efficiency~\cite{liu2021livemap}. This scenario is illustrated in Fig. \ref{fig:smartmobility}. 

Another critical transportation application is smart parking. VaaS naturally enhances the effective utilization of parking systems by capturing real-time information in parking lots and inferring useful knowledge (e.g., parking space availability) based on onboard algorithms. What is more, the VaaS framework can monitor and predict the incoming and outgoing traffic of parking lots, thereby making recommendations on parking and navigation guides for drivers. In a nutshell, vehicles can form a mesh of SCCSI service networks to update the transport information of a city to inform visitors or citizens.

Finally, we could not end this subsection without discussing traffic control in a smart city. It has been demonstrated that even a small proportion of autonomous vehicles could help regulate traffic flow and curtail potential congestion in hybrid traffic flow (consisting of autonomous vehicles and human-driven vehicles)
\cite{wu2017emergent,wu2018joint,wu2018stabilizing,wu2021flow}. It has also been observed that piecewise constant policies could mitigate congestion. All these nice features come from transportation analysis and experiments. Considering the penetration of intelligent vehicles, particularly the controllable SCCSI-empowered vehicles in VaaS, more information about the vehicular traffic flow and road conditions are made available timely, how to leverage real-time traffic flow information to design more effective traffic flow control will become a very interesting research problem. VaaS not only has sufficient resources for collecting and analyzing real-time traffic information in a city, but also incentivizes a sufficient number of SCCSI-empowered vehicles to excise the distributed traffic control to regulate the traffic. This vision falls into flow control of hybrid traffic, which is depicted in Fig. \ref{fig:smartmobility}.

\subsection{Smart and Connected Health} 
Improving residents’ healthcare is another important design goal of a smart city. As wearable health devices have become popular, people can monitor their health state and their well-being proactively. However, this is effective only within their confines (e.g., homes or communities). When residents are on the move, they may lose the always-connected mode to receive their connected healthcare monitoring. With the envisioned VaaS, healthcare machine learning tools can run data analytics for patients or people injured on the road (either passengers or pedestrians) and offer them health recommendations or contact ambulances/hospitals immediately. Even further, elderly people could be monitored in public for possible signs or gait changes of potential disease episodes (e.g., heart attacks or heat stroke) so that proactive actions can be taken. Medical doctors or health caregivers, who are at the right time on the spot, could be immediately reached over VaaS for emergency situations. While privacy is one concern here, privacy-preserving mechanisms can always be adopted to obscure pedestrians' identities~\cite{senior2005enabling}. To get a step further, residents with chronic diseases can continue collecting their vital signals even when they get into their cars or public vehicles to run their errands. In a sense, a patient can opt for being monitored 24/7 without any stoppage, which is critical for holistic monitoring and early detection of health deterioration. 

VaaS can also help improve public health. In the COVID-19 pandemic era or flu seasons, human mobility data can assist city authorities in managing the epidemic and guide citizens to act accordingly to minimize infection risks~\cite{hao2020understanding,chen2021age}. To achieve this goal, SCCSI-empowered vehicles can naturally collect pedestrian mobility data from the onboard sensors, evaluate the infection risks of different sites, and disseminate guidance/recommendations to citizens to curb the spread of an infectious virus.

\subsection{Smart Living}
Our VaaS can also provide value-added services, such as dining recommendations, to tourists and citizens, and more importantly for the city to attract more business. Back to Alice's example, to avoid waiting in lines for hours for a great meal, it will be beneficial if a smart city provides real-time information about the waiting times at certain restaurants just at users’ fingertips. To provide the services, SCCSI-empowered vehicles can capture visual images (e.g., photos of restaurants) when passing by, conduct analytics on the waiting time, and disseminate the information/recommendations to interested citizens. If a specific restaurant of interest is too crowded, restaurants with similar food and faster availability can be recommended and pushed to those who are interested. 

Another value-added service is the ad hoc community formation over VaaS. Since participating SCCSI-empowered vehicles are potentially entrusted over the smart city service network, they could form or search for an ad hoc community in which they could engage activities of common interest, such as jogging over a well-known trail, square dancing at certain spots, and singing at certain hours. Given that people tend to engage in these activities with the ones in the neighborhood, vehicles become the perfect media to disseminate such information locally. In this way, those who share similar interests could enjoy living and aging together. More importantly, they do not need to preset the time or place and could organize an event promptly over such a trustworthy network.

\section{From VANETs to VaaS: What Upgrades Do We Need?\label{limitations}}
In our VaaS, the proposed SCCSI service network can support pervasive smart city applications as illustrated above. While it is thrilling and disruptive, it is difficult to directly apply the existing VANET solutions to fulfill this vision. In what follows, we examine the limitations of current VANET solutions, which enables us to identify the needed upgrades.


\subsection{Co-Existence with Primary Services: the Spectrum Perspective}
The first issue that comes with VaaS is spectrum usage. Although VANETs do address spectrum efficiency problems in their design, they typically address the spectrum efficiency for their own licensed bands in distributed short-range communications (DSRC) or cellular vehicle-to-everything (C-V2X)~\cite{araniti2013lte,chen2022vehicular}, and all these efforts primarily address communications services for spectrum efficiency and driving safety. When applying them to VaaS, there arise two problems. First, VaaS intends to serve as a crowdsourcing platform to handle a very broad range of services, such as offloading IoT traffic in smart cities~\cite{ding2019beef,ding2018smart,ding2018intelligent}. For this reason, we must address the co-existence of multiple-typed services with different priorities such that the delay-tolerant and non-safety-related services, particularly, the ``crowdsourced services'' or secondary services, should not impact the provisioning of mission-critical vehicular applications such as autonomous driving. Second, DSRC and C-V2X will not provide ultimate solutions to the network traffic congestion problems. This is because wireless traffic, including mobile traffic, is still exponentially increasing \cite{aijaz2013survey}. According to the Shannon's capacity theorem, no matter how much bandwidth to be allocated, we will hit the spectrum ceiling sooner or later, which motivates us to always seek additional spectrum resources whenever possible. In fact, although we can always utilize high-frequency bands such as millimeter wave (mmWave)~\cite{ding2021context} and even terahertz band due to the huge bandwidth in the range, such a band is only suitable for short-range communications and susceptible to mobility and obstacle blockage, resulting in high channel variability. The situation becomes worse under vehicular environments due to the high mobility of vehicles and complicated traffic conditions, posing great challenges to the transportation of large data chunks over VANETs.

To support both primary and secondary services, the cognitive radio (CR) technology is a natural solution, which can allow secondary users to opportunistically use idle licensed spectrums as long as such a use does not significantly affect the service quality of primary users \cite{ding2017cognitive, pan2014when, pan2012spectrumclouds}. In this regard, the delay-sensitive services, such as autonomous driving applications, can be provided over the licensed bands while secondary or crowdsourced services are supported only when there is spectrum vacancy, thereby significantly enhancing spectrum utilization and provisioning differentiated QoS guarantees. We will elaborate more on spectrum management in the architectural overview in Section \ref{architecture}.


\subsection{Co-Existence with Primary Services: the Computing Perspective}
Although vehicular edge computing has attracted great attention, the current research and industrial efforts attempt to utilize roadside edge computing infrastructure to facilitate vehicles on the road. On the contrary, VaaS reverses the process to employ the underutilized resources on vehicles for public interest. This line of research is still in its fancy. With the advent of cloud/edge computing, there are some papers proposing to employ vehicles as edge computing servers. Although the concept of vehicular cloud computing has been suggested (see~\cite{abuelela2010taking,eltoweissy2010towards,olariu2011taking}), some existing works stand only at the conceptual level, leveraging cars along streets, in parking lots, and/or on the move as computing servers~\cite{liao2019fog,lee2014vehicular,gerla2014internet,lu2014connected,hou2016vehicular,hou2017wireless,boukerche2018vehicular, hou2018novel}. Some other works study task offloading to vehicles ~\cite{ghazizadeh2014scheduling,feng2017ave,zhu2018fog,sun2019adaptive,su2018distributed,zhou2019computation,pang2019spath}.

However, one pivotal problem remaining largely uncharted is how to address the co-existence of secondary computing services (e.g., the services provided by vehicles to others) with primary vehicular services (e.g., autonomous driving). Since vehicles provide services to others only when their resources sit idle, the uncertainty of computing capabilities must be considered, modeled, and predicted in order to fulfill our vision without degrading the performance of primary vehicular services, particularly safety-related operations. What is more, vehicular mobility adds another dimension of resource uncertainty as the servers (i.e., vehicular computing capability on vehicles) may come and go, making computing resource prediction and management more challenging in VaaS. To provide certain QoS guarantees, a centralized controller with a resource database is necessary, which maintains a dynamic computing resource map, just like the spectrum map in cognitive radio systems~\cite{ding2017cognitive,yilmaz2013radio}!

\subsection{Integrated SCCSI Design}
Traditional research problems in VANETs, such as data forwarding and routing, are long-standing and have been extensively studied\cite{karagiannis2011vehicular}. However, these works focus on information data delivery without taking sensing, storage, computing, and intelligence into account. With the fast convergence of communications and computing in 5G+, it is important to employ vehicles as not only communication nodes but also powerful edge nodes for sensing, storage, computing, and learning. With this in mind, the design for vehicular networks becomes ``service-oriented'' and must consider integrated SCCSI design. First, due to the nature of the wireless medium, the sensing, computing, or storage devices (all can be vehicles in VaaS) share the same spectrum resources. The scheduling and utilization of network-wide sensing, computing, and storage resources cannot be effective without an appropriate design of communication networks to move data around. Second, the emerging smart city applications can involve SCCSI resource provisioning in an end-to-end (e2e) fashion, where each of the components will contribute to the QoS metric, such as e2e latency or accuracy. For instance, for edge inference tasks, inference accuracy is one of the most important QoS metrics. More sensing data can generally enhance inference accuracy, but meanwhile, increase communication and computing overhead. An intuitive solution is to only transmit the ``important'' sensing data while discarding the redundant or insignificant data without noticeably affecting the learning/inference accuracy. This simple instance demonstrates that SCCSI resources must be jointly coordinated, and the tradeoff should be carefully exploited to optimize the service-oriented metrics under resource constraints. Unfortunately, the integrated SCCSI design has yet to be carefully studied for vehicular networks. For example, although vehicular crowdsensing has been investigated~\cite{xu2019ilocus}, the prior schemes merely focus on data collection without considering the e2e design to incorporate communications and computing for the sensed data.

\subsection{Sustainable Incentive Design}
Compared with 5G+ systems, the rollout of VaaS requires much less investment in infrastructure installment as only partial infrastructure (e.g., collocating edge servers) is needed. Besides, SCCSI capability in vehicles can be purchased by the vehicle owners, who thus could share the cost of VaaS! Nevertheless, it does demand operational costs due to the resource/energy consumption of vehicles, which must be compensated by the system via rewards (either monetary or credits) in order to attract vehicles' participation to contribute their resources for smart city services or operations. Although 5G+ infrastructure also incurs huge operational and energy costs, such cost information is known to the system operators. In contrast, in VaaS, resource consumption occurs on the vehicle side, which is not directly known by the system operators, leaving the space for vehicles to ask for rewards much higher than their true expenditure, potentially defeating the purposes of economic attractiveness of VaaS. For this reason, appropriate incentive design, perhaps based on game theory, tailored for VaaS, is in dire need to ensure the robustness to market manipulation, making it sustainable and economically attractive to communications and computing industrial stakeholders or city authorities. 

\section{The Architecture of the SCCSI Service Network\label{architecture}}
Driven by the above upgrade needs, in this thrust, we will elaborate on our proposed architecture of SCCSI service networks.

\begin{figure*}
\centering
\includegraphics[width=0.7\textwidth]{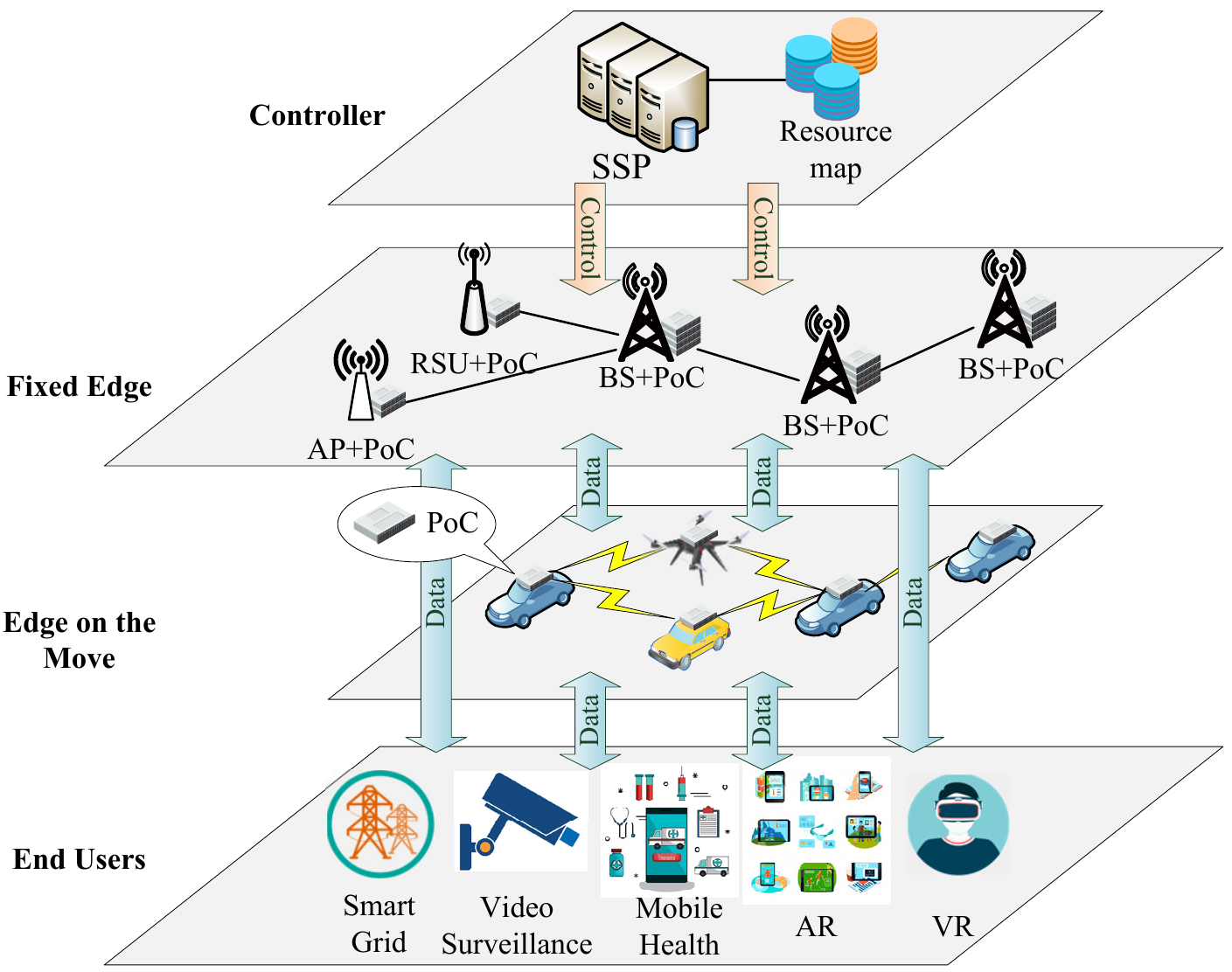}
    \caption{The overall system architecture of the SCCSI service network. VaaS attempts to tightly integrate both fixed (e.g., 5G+) and mobile (e.g., vehicular) infrastructure to provide services in smart cities. }\label{fig:system_model}
\vspace*{-0.1in}
\end{figure*}

\subsection{An Overview\label{Overview}}
Our design objective is to leverage the SCCSI capabilities of vehicles to provide SCCSI services to almost all kinds of IoT applications (vehicular or non-vehicular applications) in smart cities. As alluded earlier, the SCCSI service network harvests the SCCSI resources that are inherently uncertain and subject to vehicles' primary use. To better leverage the resource and mobility opportunity, a supporting network architecture is essential, which must 1) address the co-existence of primary and secondary services, 2) support integrated SCCSI service provisioning with dynamic resources, 3) incorporate an incentive platform to stimulate participation, and 4) ensure the seamless integration with fixed infrastructure, especially with 5G+ networks.

We envision that our SCCSI service network satisfies the aforementioned requirements. As shown in Figure~\ref{fig:system_model}, an SCCSI service network consist of \emph{end users}, \emph{points of connections (PoCs)}, and a \emph{secondary service provider (SSP)}. Moreover, to tackle vehicular mobility and the uncertainty of resource availability, it is well recognized that centralized intelligence could help enhance resource utilization and provide certain QoS guarantees (following the design principle of software-defined networking (SDN) and 5G+ virtualization). The proposed SCCSI service network features centralized intelligence with control/data plane decoupling (which is aligned well to 5G+ systems), cognitive radio (CR) enabled networking, and an incentive platform. We elaborate on these in details next. A comparison between traditional VANETs and the SCCSI service network is provided in Table \ref{comparison}.

\begin{table}
\centering
\caption{Comparisons between traditional VANETs and the SCCSI service network.}
\begin{tabular}{|l|l|l|} 
\hline
 &  \makecell{\textbf{Traditional VANETs}} &  \makecell{\textbf{SCCSI Service Network}}  \\ 
\hline
\makecell{\textbf{End users}} & \makecell{Vehicles only} & \makecell{Various IoT devices, \\not limited to vehicles}  \\ 
\hline
\makecell{\textbf{Role of vehicles}} & \makecell{End users mainly} & \makecell{Resource suppliers and\\end users}  \\
\hline
\makecell{\textbf{Resource}\\ \textbf{Management}} & \makecell{Separate design\\ of SCCSI} & \makecell{Joint design\\ of SCCSI}  \\ 
\hline
\makecell{\textbf{Communications}} & \makecell{C-V2X and/or\\ DSRC mainly} & \makecell{C-V2X and/or DSRC\\+ Cognitive radios}  \\
\hline
\end{tabular}\label{comparison}
\end{table}

\subsection{SCCSI Service Network Components}
In this subsection, we introduce the important components of our SCCSI service network. 

\begin{figure}
\centering
\includegraphics[width=0.45\textwidth]{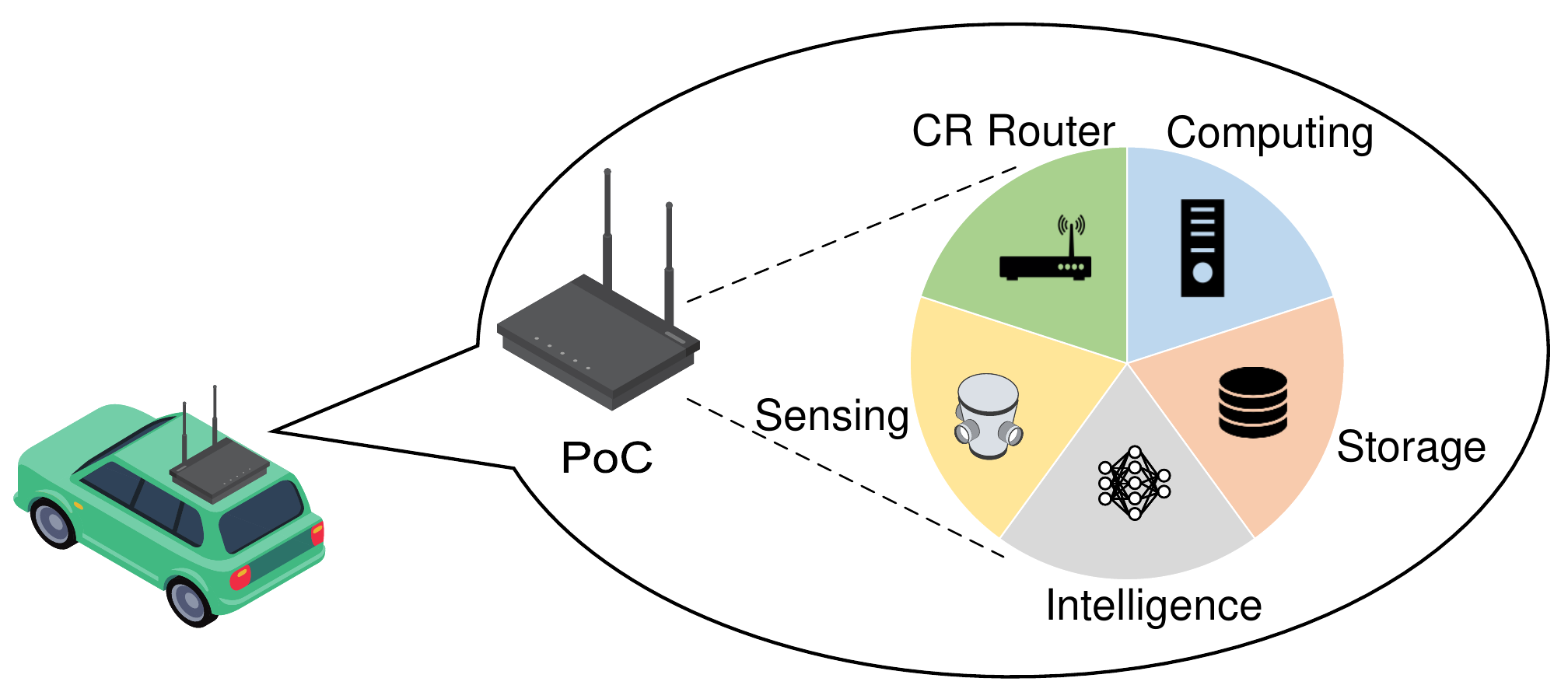}
    \caption{The illustration of PoC, which is a network connecting device equipped with SCCSI capabilities.}\label{fig:PoC}
\vspace*{-0.1in}
\end{figure}

\textbf{Point of Connection (PoC):}
We have introduced a concept called {\em Point of Connection},  simply PoC, in our previous work \cite{ding2019beef}. As shown in Fig. \ref{fig:PoC}, it is a network connecting device equipped with SCCSI capabilities. Specifically, we assume that PoCs consist of customized sensing capabilities (e.g., cameras, radars, and LiDARs), customized communication capabilities (e.g., software-defined radios or cognitive radios, which can be tuned to any end device's radio interface to directly communicate), powerful computing resources, and sufficient storage resources. We envision that VaaS installs customized PoCs at strategic locations to beef up their SCCSI capabilities. On the one hand, VaaS can leverage PoCs to build the partial fixed infrastructure as we alluded earlier. For example, a PoC can be installed at a (small) base station in cellular systems, an access point (AP) in WiFi systems, or a cognitive radio (CR) access point in CR networks \cite{ding2017cognitive}. On the other hand, PoCs can be used to empower the SCCSI capabilities in mobile infrastructure, which is the driving idea of VaaS. Specifically, a PoC can be installed in vehicles, such as city-owned buses and subways, police cars, city service trucks, taxis, private passenger cars, and UAVs. In this way, PoCs form a high-speed backhaul service network to perceive environments, handle data traffic, perform computing tasks, or temporarily store data. In particular, the vehicle-mounted PoCs can either refer to the set of equipment embedded in autonomous vehicles (which naturally has powerful SCCSI capabilities!), or can be upgraded or purchased by vehicle owners because they desire better in-car user experiences or even because they want to earn extra money from the system (recalling that they will be rewarded if providing services to others). \textit{Throughout this paper, when mentioning vehicles that provide services, we implicitly refer to vehicles equipped with such PoCs.}

Based on different criteria, we can classify PoCs into the following categories. On the one hand, PoCs can be either deployed by SSP or privately owned by citizens. For the latter case, sufficient incentives must be created for their PoCs in order to crowdsource their underutilized resources for service provisioning. On the other hand, PoCs can serve as either opportunistic edge nodes (e.g., private cars or cargo UAVs) or controllable edge nodes~\cite{li2020energy}. For opportunistic PoCs, their high-priority task is to travel to their destinations, while only providing services when their resources are underutilized at the right time and right spot. To make use of these resources opportunistically, the system must handle their mobility uncertainty and resource availability, which may change over time in response to varying traffic conditions or primary service demands. For controllable PoCs, they can be either fixed infrastructure or vehicles (including ground vehicles and unmanned aerial vehicles or UAVs) dispatched/controlled/incentivized by the network operators (e.g., SSP running a smart city) to help handle the bursty service requests with carefully designed trajectories/speeds.

When sufficiently many vehicles are PoC-equipped, the fleet of roaming vehicles may provide plenty of mobile sensing, communications, and computing resources all over the city, enough to push the SCCSI services closer to end users at the edge as vehicles tend to have a higher probability of reaching the curb. By getting closer to end users, short-range transmissions with low transmit power can be carried out between end users and PoCs for data delivery and/or computing, which can remarkably improve spectrum utilization, offload cellular data traffic~\cite{ding2017cognitive}, and satisfy ubiquitous communications/computing needs without requiring extravagant upgrade on existing infrastructure. 




\textbf{End Users:} 
In traditional vehicular networks, end users are generally vehicles or vehicular occupants, and the main focus of the related research works is to support vehicular communications in order to ensure road safety, improve traffic efficiency, and provide infotainment services to passengers. The end users in VaaS, however, comprise almost all kinds of IoT devices in either urban or suburban areas, such as vehicles, smartphones, VR/AR devices, street cameras, and traffic/environmental/health sensors. Vehicles or vehicular occupants are just one kind of users among them. Note that vehicles can be either end users acquiring services, PoCs providing services to diverse IoT devices, or serve as both at the same time. 

\textbf{Secondary Service Provider (SSP):} One salient feature of VaaS is the leverage of vehicular mobility and spatial variations in spectrum availability (spectrum mobility) to exploit joint communication and computing opportunities. Due to the uncertainties, a certain level of centralized control may more effectively provision the spatio-temporal varying resources by proactively gathering sufficient information. For example, the centralized intelligence can enable fast handshake between transmitters and receivers not to miss pending opportunities~\cite{ding2018smart,ding2017cognitive}, or quickly set up multi-hop transmissions with certain QoS guarantees before opportunities pass. Following the design principle of SDN, we introduce a \emph{secondary service provider (SSP)} into VaaS to manage control signaling. As we have suggested in~\cite{ding2017cognitive}, SSP could be an independent service provider that explores a new service business opportunity, or an existing service operator (e.g., a 5G+ cellular operator) that harvests in-network resources to address its communications and computing needs, or even a city authorized organization who manages the information and communications system for a smart city. No matter what it is, it must have its own reliable spectrum resource, called {\em basic bands} (\eg, cellular bands if SSP is a cellular operator), which can be used to provide common control signaling (CCS) to manage the network operations (pretty much the same as in 5G networks). It has been demonstrated that it is always much more effective to advocate the decoupling of control and user plane (C/U-plane decoupling \cite{yan2015control,song2016millimeter}) for high-speed communications to simplify the switching fabrics in user data plane. 

As a remark, the term ``secondary service provider'' stems from cognitive radio networks~\cite{ding2017cognitive}, where the communications services will be opportunistically provided to secondary users only when their interference does not affect the QoS of primary users. Analogous to cognitive radio networks, VaaS harvests the resources on vehicles and spectrum bands for the common good only if the resources are underutilized, thereby minimizing the impact on the primary services the drivers or vehicle occupants demand. We call these services ``secondary services'', as vehicles generally also have their primary tasks (e.g., self-driving decisions) to do. As argued in \cite{ding2017cognitive}, stochastic opportunities can only be leveraged more effectively with a centralized entity, the one that holds a global picture. SSP proactively collects information about tempo-spatial SCCSI capabilities and makes the best use of them opportunistically for service provisioning in VaaS, which echos the commonly spoken statement ``be prepared for the opportunities''. 


\subsection{Functional Design}
For effective control and resource management, we consider the following basic operations that should be enabled by SSP.

\subsubsection{CR-Enabled Data Transmissions} 
To support the co-existence of primary and secondary services in VaaS, we can resort to CR technologies. As we demonstrated in Section \ref{limitations}, the existing solutions (DSRC or C-V2X) face significant challenges when supporting diverse IoT services due to spectrum scarcity, calling for the use of cognitive radio technologies. In VANETs, cognitive radios and harvested spectrum resources have been proposed~\cite{zhou2016toward}. Unfortunately, each CR-enabled vehicle still accesses these harvested spectrum resources individually and passively without good coordination for network-wide data transportation. If devices operate on their own in terms of spectrum harvesting and service provisioning, both spectrum efficiency and energy efficiency will be low. The licensed spectrum may be highly dynamic, depending on the activities of primary users. A secondary user intending to use a licensed band to establish a communication session has to conduct spectrum sensing to harvest appropriate licensed bands for opportunistic use, resulting in low energy efficiency and slow handshake. Besides, secondary users may have to compete for an idle channel after spectrum sensing, i.e., they have to share already rare opportunity, and thus, without effective coordination, the spectrum efficiency will be low~\cite{ding2017cognitive,ding2018session}. 

In VaaS, PoCs are expected to be equipped with CR capabilities, forming a web of CR routers. Based on this idea, in the last few years, we have advocated network-wide architectural design to take full advantage of CR technologies, where vehicles serve as mobile CR routers to proactively perform spectrum sensing for all secondary users and transport data between devices and data networks~\cite{ding2018smart,ding2019beef}. In our prior research, we demonstrate that coordinated sets of well-designed CR routers could more effectively manage idle licensed spectrums by bringing network services closer to end users (\eg., SUs)~\cite{ding2017cognitive}. In VaaS, the CR routers can be in ubiquitous vehicles, either parked or on the move. Under this design, locally available spectrum with uncertainty could be better utilized with higher spectrum efficiency by centralized management. Besides, by considering multi-hop transmissions and/or store-carry-forward transmissions, end users, vehicles, and roadside infrastructure can communicate with each other with significantly shorter distances. As a result, the energy efficiency can be considerably reduced while mitigating potential interference among their transmissions. It is noted that, in addition to the CR capabilities of vehicles, this solution does not impose specific requirements on the radio interfaces of IoT devices or pedestrians, since vehicles can configure their CR interfaces to inter-operate with devices from heterogeneous systems (e.g., cellular, WiFi systems, or Zigbee) through the interfaces that these devices normally use.

\subsubsection{SCCSI Resource Map}
To harvest the resources from vehicles, SSP has to maintain a real-time SCCSI resource map in its database, as illustrated in Fig. \ref{fig:system_model}. Considering the dynamics of resource availability, such a map is similar to a spectrum map in cognitive radio systems~\cite{yilmaz2013radio,ding2017cognitive}, yet with multi-dimensional resources. With vehicles roaming around a city, they can easily interact with IoT devices in the city when they get close, collecting information from and pushing services to the curb! One important issue in SCCSI service provisioning is the collection and distribution of information about SCCSI resources and their availability. Vehicles could collect the required information to help SSP develop a (temperature) color map for resource availability and service demands. For example, selected vehicles could be used as spectrum sensors to detect spectrum availability and also serve as computing load monitors to collect computing workloads on edge servers. Moreover, participating vehicles could provide their computing capability information, storage availability, and machine learning (ML) tools to the nearby mobile edge nodes. In this way, the real-time resource ``color map'' could be updated and made available for use. 

As a remark, in addition to SCCSI resource availability, with machine learning tools and multi-modal sensors on vehicles, non-physical sensing such as community commotion sensing, neighborhood disturbance sensing, noise level sensing, and/or air pollution sensing can also be collected and integrated into this map, which will be useful for smart living and smart healthcare. 

\begin{figure}
\centering
\includegraphics[width=0.45\textwidth]{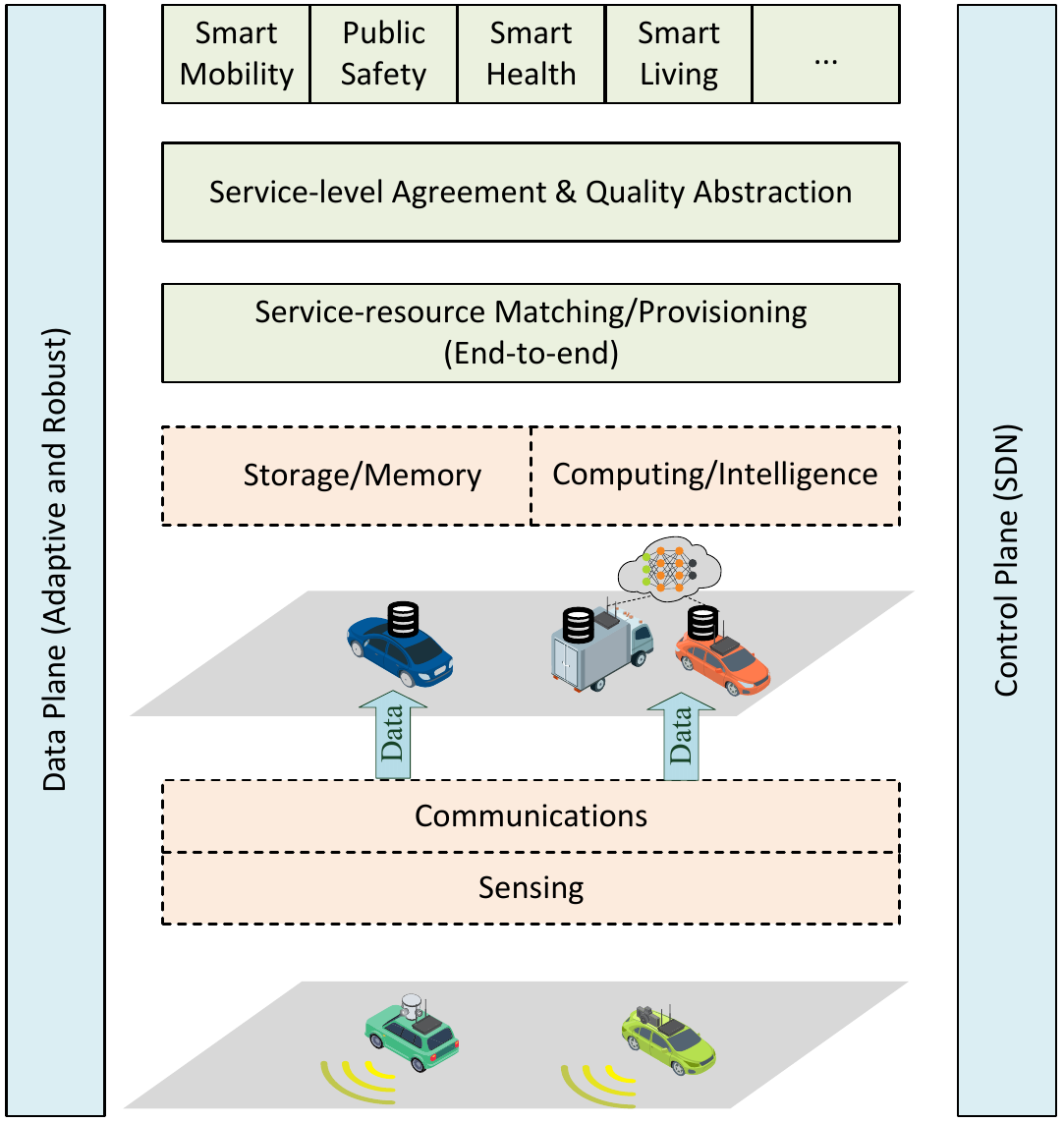}
    \caption{The integrated design for SCCSI services in VaaS.  
    }\label{fig:SCCSI}
\vspace*{-0.1in}
\end{figure}

\subsubsection{SDN-based Network Management} To acquire a global view of VaaS, SSP may select a subset of fixed PoCs to form a control signaling network to manage VaaS by following the SDN design philosophy~\cite{kreutz2015software} or Control/User Data (C/U) decoupling approach~\cite{song2017control,yan2015control}. 
VaaS is composed of a variety of heterogeneous PoCs and devices. To effectively coordinate the heterogeneous network and execute high-level network policies, SDN decouples network control functionalities from the underlying devices and enables logical centralization of network control. The illustration of the SDN-based SCCSI integration is shown in Fig. \ref{fig:SCCSI}. Except for the remaining data forwarding operations, the devices of different kinds do not need to be aware of  various protocols but only need to receive control policies from SSP, where the control signaling is carried out over the reliable basic bands (the control plane or C-plane). In this way, we can significantly reduce the complexity of network management and improve the reconfigurability and interoperability of the SCCSI service network. 

SDN can not only make it much easier to configure/reconfigure VaaS with customized and adaptive control policies, but also offer QoS provisioning guarantees by making use of the global knowledge at the central controller (i.e., SSP). Based on the statistical maps of SCCSI resource availability, SSP then solves formulated optimization problems, comes up with control strategies, sends the control policies to PoCs, and provides the SCCSI services to the end users at the edge.


\subsubsection{Incentive Platform} 
When providing services to others, vehicles have to consume their own resources, such as battery, CPU/GPU, spectrum, and storage. Therefore, it is necessary to compensate for their services. Towards this end, SSP should establish a reward platform with well-designed pricing policies to reward them. For example, SSP can employ either reverse auction or double auction~\cite{chen2022from} to draw the ask prices from vehicles, which serve as sellers to sell their provisioned services.

\subsection{The Benefits of VaaS\label{Benefits}}
To provide a better interpretation of our proposed VaaS framework, 
we will examine its benefits and how it addresses the limitations of traditional VANETs in Section \ref{limitations}.

\subsubsection{Support Co-existence of Primary and Secondary Services} Based on CR technologies and the real-time SCCSI resource map, SSP can judiciously harvest network-wide underutilized SCCSI resources for secondary service provisioning without affecting the primary or mission-critical services. 

On the one hand, ultra-reliable low latency communication (URLLC) can be supported by VaaS. To decrease the service latency for emerging services, it is better to support services as locally as possible. In VaaS, vehicles supply more ubiquitous SCCSI resources than ever before, forming an ultra-dense service mesh network. End users, therefore, can acquire a variety of services with fast response at the network edge. One may argue that the mobility of vehicles and the uncertainty of resources in VaaS contradict to the ultra-reliability requirements. However, it is noted that parked vehicles, which generally stay for hours, constitute a significant proportion of VaaS. Therefore, ultra-reliable services can be assigned to parked vehicles with resources reserved for the next period of time (based on mutual agreement), whereas non-critical services can be assigned to moving vehicles, thereby achieving differentiated QoS guarantees. On the other hand, the proposed VaaS can offload delay-tolerant services via opportunistic resource harvesting. In particular, video traffic accounts for 65.93\% of the Internet traffic in H1 2022~\cite{Sandvine2023The}, a large proportion of which is tolerable to latency. Offloading delay-tolerant traffic plays a critical role in saving cellular bandwidth for more delay-sensitive services. Unlike the previous mode for delay-sensitive service provisioning, SSP can take advantage of the opportunistic mobility of vehicles, their intermittent network connectivity, and their temporarily idle spectrum and computing resources to handle the delay-tolerant service requests without much affecting the primary use of resources or high-priority services.

\subsubsection{Promote the Convergence of SCCSI} 
While multi-access edge computing (MEC) is an indispensable component in the 5G+, the deployment of MEC is largely hindered by the installment cost. The SCCSI-empowered vehicles fill in this gap nicely and provide ubiquitous computing resources, leading to the true convergence of SCCSI. In addition, the proposed centralized controller collects network-wide SCCSI information and conducts network optimization based on SDN, making it feasible to implement algorithms by considering service-oriented metrics (e.g., end-to-end latency) and multi-dimensional resource provisioning.

\subsubsection{Improve Spectrum Efficiency and Energy Efficiency} VaaS will greatly count on short-range V2I and V2V communications, thus enhancing spectrum efficiency. By leveraging the omnipresence of vehicles, IoT devices and relaying vehicles can form a dynamic mesh network to deliver data via multi-hop communications. Compared with direct transmissions to destinations through a long distance (e.g., direct cellular transmissions to BSs), this approach allows transmitters to lower their transmit powers, thereby causing less interference to others.

This paradigm shift also improves energy efficiency in the sense that lower transmit powers can be used due to the reduced propagation distance and less destructive wireless impairments. Due to higher frequency bands, larger traffic loads, better spatial multiplexing, and always-on control signaling for information collection, 5G+ consumes much more energy than 4G within a similar coverage~\cite{lopez2022survey}. By exploiting multi-hop relaying and opportunistic mobility, VaaS can effectively reduce the energy consumption on both end users and radio access networks (RANs), which is aligned well to the sidelink transmissions advocated by 5G+ standards~\cite{garcia2021tutorial}. End devices can communicate with vehicles when they are getting close, instead of directly transmitting to potentially remote base stations by increasing their transmit powers. Similarly, base stations can push contents to nearby vehicles, which then serve as relays to distribute contents to intended users in close proximity. 

\subsubsection{Extend Service Coverage} While data rate can be boosted by making use of high-frequency bands, it is challenging to support large coverage of high-frequency transmissions, e.g., mmWave transmissions, due to the severe path attenuation~\cite{zhang2020augmenting}. Vehicles can come to rescue because they can form a multi-hop relaying network to extend the service coverage of cellular networks while ensuring high e2e throughput. Besides boosting service experience for urban users, VaaS can also deliver services to remote areas out of cellular coverage or with poor cellular connectivity, bridging the ``digital divide'' and making 5G+ services accessible to everyone, including those in remote and poor communities~\cite{lappalainen2020bridging}. For example, vehicles can form relay networks or data mules to support data exchange between remote IoT devices and access points~\cite{yaacoub2020secure}, or serve as computing facilities to democratize edge computing/intelligence services to users in remote villages. 

\subsubsection{Make it Compatible with 5G+} The SCCSI service network has excellent compatibility with 5G+ in the sense that 5G+ operators could upgrade their networks to incorporate the proposed new features without extravagant changes. From the architectural perspective, SSP can be a 5G+ operator. Moreover, the SDN design philosophy in the SCCSI service network is aligned well to 5G+~\cite{huawei20165g}. From the communications perspective, 5G V2X communications, including UE relaying in 5G V2X standards~\cite{garcia2021tutorial}, can be easily adapted to build up the proposed service mesh network. From the computing perspective, the holistic design of communications and computing is also the development trend in 5G+~\cite{huawei20216g}. In a nutshell, the SCCSI service network exhibits excellent compatibility and can be seamlessly integrated into 5G+ mobile networks.

\section{Sensing as a Service\label{sensing}}
\begin{figure}
\centering
\includegraphics[width=0.45\textwidth]{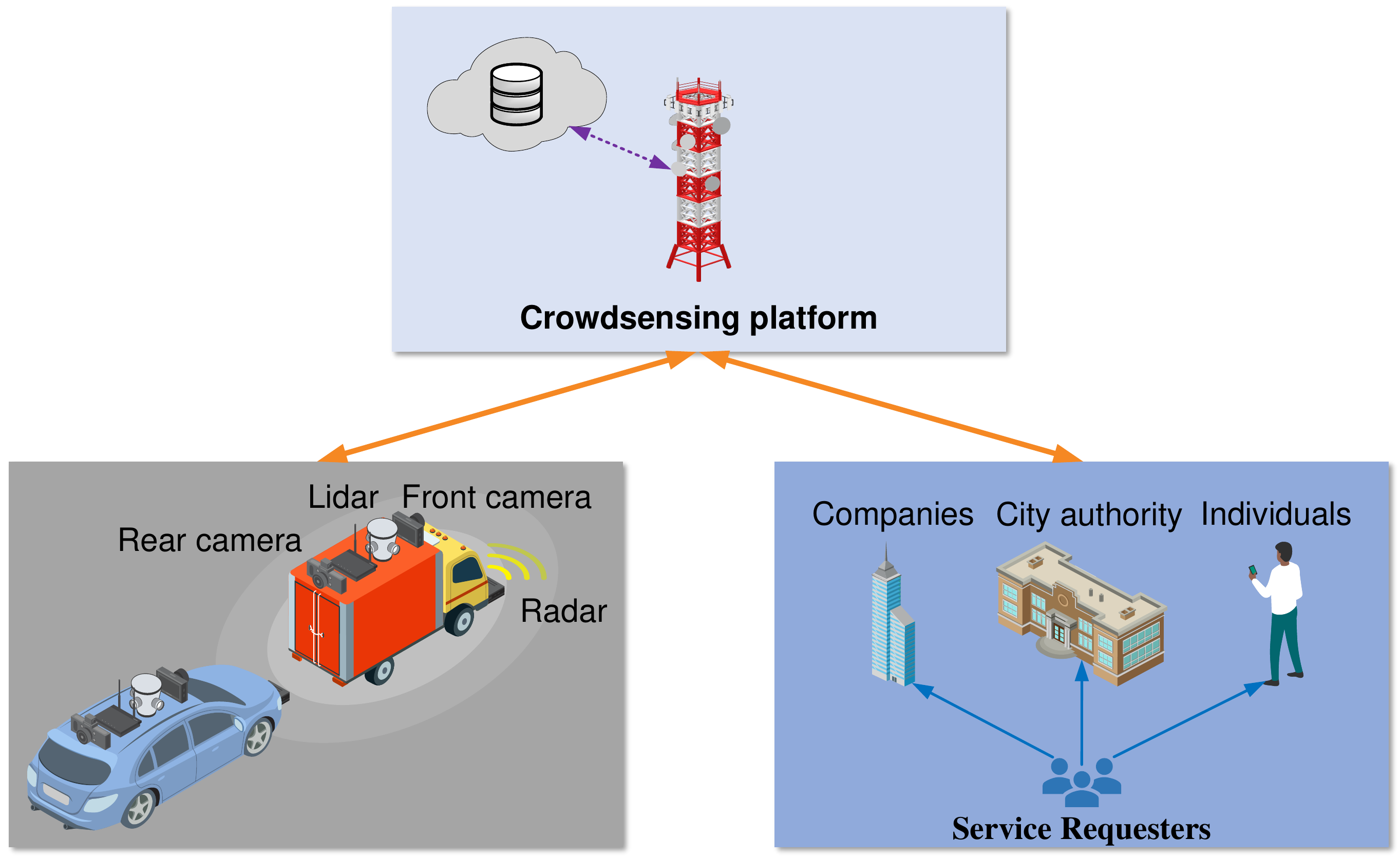}
    \caption{Crowdsensing in VaaS. In smart cities, vehicles share their sensory data with other interested parties or citizens for the common good.}\label{fig:sensing}
\vspace*{-0.1in}
\end{figure}
After illustrating the overall architecture, we next present how vehicles could provide SCCSI services in smart cities. Unlike prior surveys mainly focusing on single-dimensional resource management and related topics, we will deliver a comprehensive picture, with an emphasis on the convergence of SCCSI in a multi-dimensional view. We start with sensing.

Vehicles are being equipped with a variety of sensor devices such as high-resolution cameras, light detection and ranging (LiDAR), and radars~\cite{liu2019edge}. The multi-modal sensory data can not only be utilized by vehicles to perceive the surrounding environments for smart driving, but also be shared with other interested parties, such as city authorities, commercial companies, and citizens, forming the paradigm of ``vehicular crowdsensing''~\cite{he2015high}, as illustrated in Fig. \ref{fig:sensing}. Vehicular crowdsensing empowers a wide range of revolutionary applications, laying the foundation for the intelligent operations of smart cities. For instance, vehicular crowdsensing can enable real-time high-definition (HD) map update, which is essential for autonomous driving and robot navigation in urban areas~\cite{liu2021livemap}. Vehicular crowdsensing can also monitor and detect the abnormal events like crime progressions to enhance public safety. If machine learning tools are installed in vehicles, community commotions or emotional feeling can be detected through digital content analysis or scene analysis~\cite {cuff2008urban}. 

To enhance data collection opportunities, several aspects of vehicular sensing, such as vehicle selection, incentive design, and security and privacy, have been studied in the literature. Given the sensing and data upload costs, vehicle selection strategies must be dependent on vehicles' future trajectories, timeliness, and costs~\cite{chen2022timeliness,gao2018truthful,xu2019ilocus}. Ideally, a crowdsensing platform aims to employ vehicles at the right locations and right time while incurring the minimum costs (requiring the minimum incentives) or within budget constraints. In \cite{he2015high}, He et al. devise a vehicle recruitment scheme for vehicular crowdsensing based on predictable mobility to maximize spatial and temporal coverage. A polynomial-time greedy approximation algorithm is proposed to solve the problem with an approximation guarantee. In \cite{xu2019ilocus}, based on the observation that taxi agents typically have different goals from a crowdsourcing platform, Xu et al. combine both monetary rewards and potential ride requests as the incentives to minimize the dissimilarity between taxis' trajectories and the desired crowdsensing distributions. Furthermore, by considering game-theoretical models, some works formulate vehicle selection and pricing strategies as reverse auction mechanisms. In \cite{chen2022timeliness}, by taking travel time into consideration, we propose a timeliness-aware incentive mechanism based on reverse auction to maximize the platform utility in terms of time. This scheme is well-suitable for the case where timeliness is of importance for task completion. In \cite{gao2018truthful}, Gao et al. present a reverse auction scheme for vehicular crowdsensing by considering travel trajectory uncertainty, where each vehicle has several potential future trajectories to traverse and hence submit bids for different trajectories.

As alluded earlier, the prior schemes on vehicular crowdsensing merely focus on vehicle selection and incentive design without considering integrated SCCSI design. In particular, vehicular crowdsensing relies on intensive data communications. The multi-modal sensory data generated by vehicles offers abundant information, but at the same time, poses grand challenges to sensory data transmissions. Taking public safety applications as an example, a naive approach is to ask all vehicles in a certain region to upload visual data for detecting crimes or tracking suspects. However, it can easily overwhelm the communication systems. For this reason, SSP must identify the ``important'' data from a subset of vehicles. For model training, the features of the data must be representative and related to target scenarios. For data analytics (model inference), the data must be highly likely to contain the object of interest (i.e., a lost child or a suspect). Along this line, we can resort to cooperative perception for autonomous driving, the idea of which is to extract useful information from rich raw data to significantly reduce the communication payloads~\cite{Where2comm,wang2020v2vnet,chen2019f}. Although these works primarily target cooperative autonomous driving by enabling sharing of complementary information among vehicles, the developed schemes can be easily adapted to update real-time city-wide information, such as road traffic, crowd mobility, restaurants, and public safety, which is invaluable for improving quality of lives and facilitating city operations. In this respect, Hu \textit{et al.} proposed a communication-efficient collaborative perception framework specifically for task-oriented vehicular communications by focusing on perceptually critical areas~\cite{Where2comm}. Specifically, they propose a spatial confidence map, which reflects the spatial heterogeneity of perceptual information, empowering agents to only share spatially sparse yet perceptually critical data. Similarly, Chen \textit{et al.} developed a point cloud feature-based cooperative perception framework to overcome the constraints of network bandwidth and real-time processing \cite{chen2019f}. Drawing inspiration from task-oriented communication, Wang \textit{et al.} designed a novel collaborative approach that intelligently aggregates information from multiple nearby vehicles, enabling the observation of the same scene from different viewpoints \cite{wang2020v2vnet}. 



\section{Communications as a Service\label{communications}}
Data transportation is the core functionality of VaaS, which not only supports data traffic delivery between sources and the Internet/intended destinations, but also serves as the basis for sensing, edge storage/caching, computing, and edge intelligence. To relieve network congestion, vehicles can help in two aspects, i.e., forming multi-hop wireless backhaul or carrying delay-tolerant data. First, vehicles can take advantage of short-distance relaying communications to significantly enhance spectrum efficiency and boost system throughput by forming multi-hop wireless backhaul for various IoT devices. Second, vehicles can also utilize their mobility to carry delay-tolerant data traffic of large volumes in a ``store-carry-forward'' fashion. In this section, we elaborate these aspects in detail.

Note that employing vehicles for relays~\cite{abdelhamid2015vehicle} and mobile data carriers (e.g., data mules~\cite{shah2003data}) both are long-standing topics. We will review these topics and discuss open research issues by considering the new application scenarios under the SCCSI service network in smart cities.

\subsection{Data Transportation Based on Vehicular Multi-hop Backhauling}
Multi-hop relaying can extend communication range and improve spectral efficiency in vehicular networking, which has already attracted significant attention. In the 3GPP Release 17 for V2X items, UE relaying has been included, with possible forward compatibility for multi-hop relaying (i.e., the sidelink communications)~\cite{garcia2021tutorial}. However, extensive research works on VANETs have only tackled information delivery and routing over vehicular networks for traffic information exchange and dissemination. The idea of leveraging them as mobile backhaul for low-end IoT devices to carry data traffic has received much less attention.

Future IoT applications involve continuous data streaming between IoT devices and data networks (e.g., fixed PoC connected to the Internet) for data collection or data analytics. In response to this need, VaaS is intended to establish mobile infrastructure to provide data connections for ubiquitous IoT devices. The participating vehicles in the SCCSI service network are envisioned to be endowed with CR interfaces, which can be tuned to the bands that mobile/IoT devices use. Within dense vehicular networks, multi-hop relaying can be employed to enhance the link capacity for devices at cell edge or create links for devices out of cellular coverage. Furthermore, vehicles can also transport massive IoT data to powerful PoCs for processing/computing, thereby offloading cellular traffic. For example, in the case of AMBER Alert, SSP can coordinate massive wireless surveillance cameras to relay their video clips via a mesh of vehicles to edge computing nodes around (e.g., parked vehicles with sufficient idle resources) for video analytics without having to go through cellular base stations, thereby significantly relieving the burden on cellular systems.

In VaaS, there can be two kinds of vehicular mesh networks, which are formed by parked vehicles and moving vehicles, respectively. For stationary vehicular multi-hop networks, the vehicles can be parked along streets or in parking lots, creating chain-like or mesh-like backhaul networks, similar to conventional multi-hop backhaul networks. Besides, when vehicles are stopped at signalized intersections, they naturally form a static mesh network for a certain time period. In \cite{liu2011pva}, it has been shown by theoretical analysis and simulations that, even when a small proportion of parked vehicles participate in transmissions, the network connectivity can be improved significantly. In \cite{liu2019parking}, the routing protocol for emergency data delivery in VANETs by considering the parked vehicles has been studied. A spider-web routing protocol, which is inspired by spider webs, is proposed to reduce the average end-to-end latency based on geographic information for parking areas. The vehicular backhaul network can also be semi-stationary or even highly dynamic. When there is a traffic jam in urban areas, the slowly moving vehicles can form semi-stationary mesh networks or relatively stationary but fast moving vehicles can form platooning networks. Of course, when vehicles are moving at a relatively high speed, network management would be more challenging. Data delivery in such highly dynamic networks can be handled by position-based data routing, where the best route is chosen based on geographical locations~\cite{leontiadis2007geopps}. Along this line, SSP can coordinate data relaying by sending commands to specific locations instead of specific vehicles. In this way, any vehicle passing through the designated locations can take the responsibility of receiving/relaying/uploading data. 

To maintain multi-hop connectivity, we may need to have plenty of vehicles traversing the city, which may not be possible in the early hours or later hours. Fortunately, when there are not enough vehicles on the road, there may not exist a large number of service users, and hence the existing infrastructure (e.g., cellular systems) could have enough spectrum bandwidth to support them with reasonable quality of service (QoS) guarantee. When the number of vehicles increases, the number of content consumers tends to increase, but there are also more vehicles to form a denser relay network to boost network performance. This is exactly the salient features of the SCCSI service network to be leveraged to build an cost-effective service network for smart city services and operations!

\begin{figure*}
\centering
\includegraphics[width=0.6\textwidth]{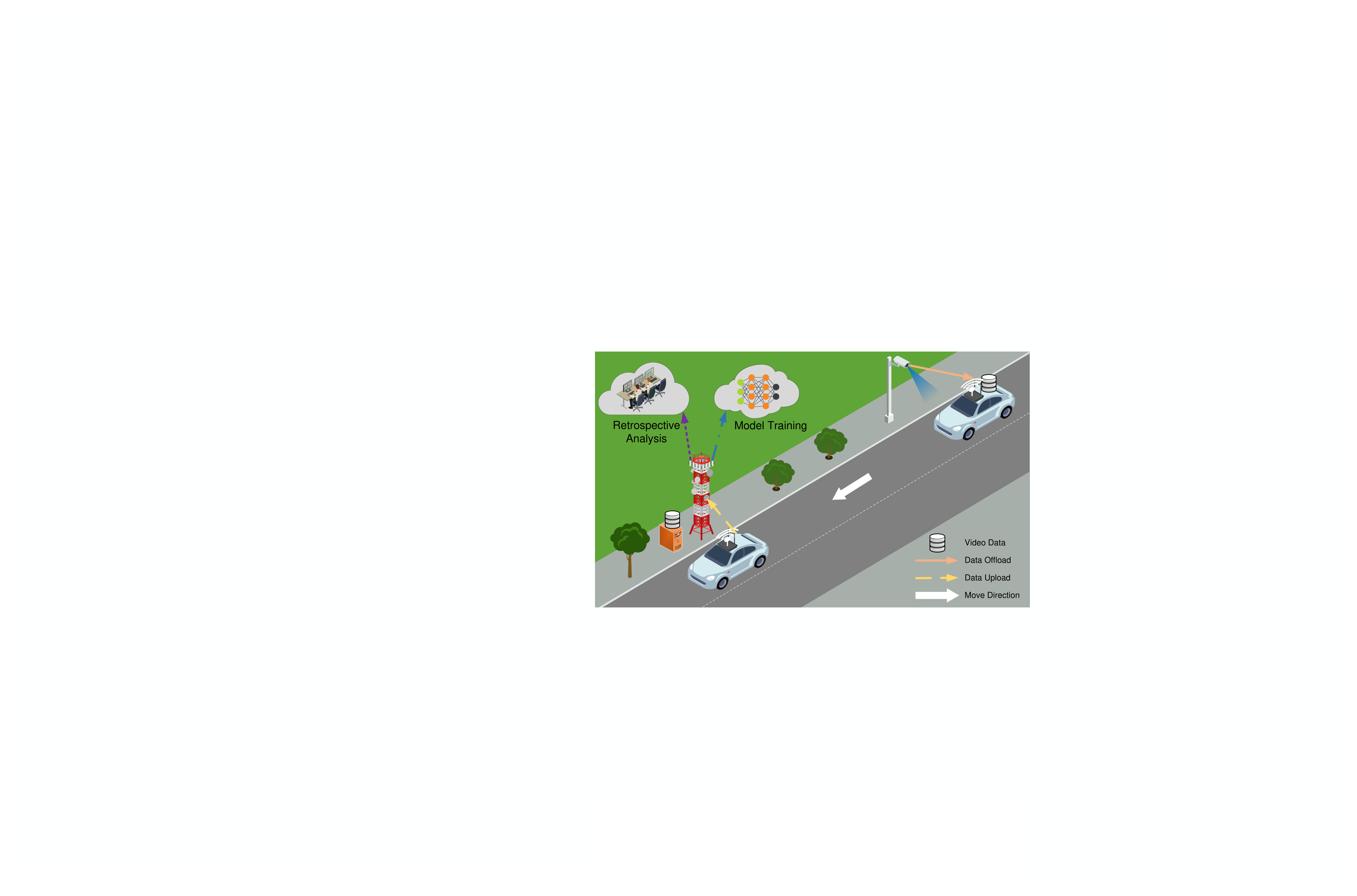}
    \caption{Data transportation based on store-carry-forward in smart cities. In this figure, a surveillance camera offloads a large chunk of video data to a vehicle, which then carries the data and uploads to a roadside access point. 
    }\label{fig:datacollection}
\vspace*{-0.1in}
\end{figure*}

\subsection{Data Transportation Based on Store-Carry-Forward}
Vehicular multi-hop backhauling can achieve the best performance on busy roads/parking lots with high-density vehicles. When the service network is sparse, however, there may be a lack of an end-to-end path from sources to destinations. Moreover, while multi-hop backhauling could provide fast and reliable data exchange between end devices and data networks, it may be unnecessary for delay-tolerant data traffic, for which our design goal is to reduce spectrum/energy consumption as much as possible. Since numerous delay-tolerant IoT applications can rely on store-carry-forward data transportation, fast data exchange, therefore, can be carried out only when vehicles approach transmitters or desired receivers (e.g., WiFi access points). As such, spectrum bands and energy can be saved for delay-sensitive services. In what follows, we illustrate two store-carry-forward modes in VaaS: content caching/pushing and data collection.

\subsubsection{Opportunistic Content Caching/Pushing} Edge caching is to prefetch popular contents from sources and place them to edge caches (e.g., base stations, wireless access points, or end-user devices) for potential future usage~\cite{bastug2014living}. For example, the trending videos in TikTok, a popular short-video platform, might be watched by numerous citizens. Also, popular AI models, such as up-to-date language models and object detection models (e.g., for autonomous driving) can also be consumed by numerous citizens. Proactively caching these contents at the wireless edge could alleviate the burden on backhaul links and backbone networks as well as significantly reduce the transmission latency, because it eliminates the need to (repeatedly) fetch the data from a remote data center or a content source upon a user request.

Although the proactive content pushing to the network edge can be surely carried out via backhaul links as the existing literature suggests~\cite{li2018survey}, it could be highly costly, especially during peak hours. In particular, for wireless edge nodes without fiber-based backhaul, such as small base stations, the proactive content pushing would be even more costly as these small base stations usually fetch the contents from macro base stations via wireless backhaul links.

Considering the delay-tolerant characteristics of edge caching used for potential future usage, VaaS becomes a perfect data transportation media to distribute such contents. The popular contents can be dumped to a vehicle from a roadside unit when they get very close. When vehicles physically move around the city, the contents can be carried, stored, and forwarded on their routes around the city. Whenever a roadside edge node is encountered and there is enough spectrum available (with the assistance of CR technologies in the SCCSI service network), the data contents can be pushed from a vehicle to a roadside edge node easily accessible for data consumers. It is noted that content pushing/caching can be conducted with the lowest priority in transmission and only when spectrum is available, particularly when they get very close. Thereby, this scheme can be called ``opportunistic content caching/pushing''. 

Edge caching based on uncertain spectrum availability over information-centric cognitive radio networks has been studied \cite{si2016spectrum}, where popular video contents are proactively disseminated and cached at the locations close to the interested users via harvested spectrum bands. However, employing vehicles to distribute contents to roadside PoCs is much more challenging due to mobility uncertainty. To reduce the design complexity, we can also employ public transit with fixed or known daily routine trajectories, such as buses, to proactively push popular contents to the network edge. Another advantage of employing buses for content pushing is that the buses will stop at bus stations for passengers to get on and off, which provides ample time for delivering large chunks of data to roadside PoCs. These ideas present intriguing problems for further investigation.

\subsubsection{Data Collection} 
Employing vehicles as “data mules” to transport data generated by IoT devices is widely considered to be an economical and effective approach for data collection~\cite{shah2003data}. Under this paradigm, vehicles are directed to pick up data from the sensors in close proximity, buffer it, and drop off the data to receivers or data networks where they could be processed, stored or consumed. Clearly, vehicle-based store-carry-forward data collection suffers from long latency. Many existing works envision this paradigm in remote areas without Internet access to transport data with low cost~\cite{yaacoub2020secure} or energy-constrained wireless sensor networks to save energy~\cite{sugihara2011path}. Nevertheless, in the smart-city context, store-carry-forward schemes can also play a crucial role in improving spectrum reuse and energy efficiency. First, tremendous amount of IoT data is delay-tolerant or at least delay-insensitive in certain time-scale. For example, if a police station intends to exploit surveillance video from wireless cameras for potential retrospective analysis or issuing traffic tickets, the services are generally not time-sensitive and can rely on vehicle-based data collection (as plotted in Fig. \ref{fig:datacollection}). The data can also be used for AI model training because model tuning/improvement is tolerable to certain latency in many applications. Second, allowing IoT devices to transmit data to passing vehicles will lead to energy saving on IoT devices because of the short-distance transmissions, which is crucial for resource-constrained devices to prolong their lifetime. Last but not least, store-carry-forward data collection enhances spectrum reuse since the transmissions can only be conducted when the data is uploaded to or dumped from vehicles in a typically short transmission range, which significantly increases spectrum reuse. As a result, base stations nearby may be able to still utilize the same spectrum bands for other transmissions without causing too much mutual interference. 

Vehicle-based data collection in smart cities does attract some attention in the past. In \cite{bonola2016opportunistic}, Bonala et al validated the feasibility and system performance of vehicle-based data collection by using real-world traces of a small taxi feet in Rome, Italy. Results suggest that even relatively small fleets, such as an average of about 120 vehicles, operating in parallel in a very large and irregular city such as Rome, can achieve an $80\%$ coverage of the downtown area in less than $24$ h. Liu et al \cite{liu2012mining} in fact made an earlier effort to use GPS data collected from taxi cabs in Shanghai for map inference, which showed very good accuracy. Clearly, if a large proportion of vehicles in a city reach a consensus to act as data mules, the quality of services in terms of coverage and latency can be significantly improved.

In contrast to existing solutions, the SCCSI service network offers significantly enhanced store-carry-forward data collection capabilities by effectively utilizing both spectrum and mobility opportunities. First, since vehicles are equipped with CR technologies, they can detect spectrum availability so that the communications between roadside IoT devices or RSUs and vehicles can be established without noticeably impacting primary services. Along this line, our prior work \cite{ding2018intelligent} proposes a spectrum-aware data transportation scheme based on store-carry-forward transmission mode to offload delay-tolerant data in smart cities. In view of the sequential data routing decisions at intersections, the store-carry-forward data delivery is modeled as a Markov Decision Process for finding the optimal data routing path. Second, due to the presence of SSP, VaaS can exploit global knowledge to gain better system performance. By taking advantage of vehicular mobility information, SSP can choose and inform transmitters and receivers before their encounter, thereby enabling fast handshake between roadside IoT devices and/or RSUs and vehicles to better utilize the short encounter opportunity. 

\section{Computing as a Service\label{computing}}
To support intelligent and computing services on resource-constrained IoT devices, edge computing has gained significant attention~\cite{deng2022actions,kuang2019partial}. By deploying computing resources at the wireless edge, end devices can obtain computing results by offloading their computing tasks to the network edge instead of running compute-intensive applications locally. In 5G+ MEC, this is achieved by placing servers at or close to base stations or radio access points. Yet, powerful computing servers are expensive, thus calling for an alternative/complementary solution. Since vehicles, either on the roads or in the parking lots or along the curbs, can form a dense mesh network of edge computing nodes, SSP can aggregate their resources to provide services to ubiquitous IoT devices in smart cities. 

The concept of vehicular cloud computing was suggested more than a decade ago \cite{abuelela2010taking,eltoweissy2010towards,olariu2011taking}, referring to ``a group of largely autonomous vehicles whose corporate computing, sensing, communication and physical resources can be coordinated and dynamically allocated to authorized users''~\cite{eltoweissy2010towards}. With the advent of edge computing, vehicular edge/fog computing has also been proposed\cite{hou2016vehicular,mukherjee2018survey,liao2019fog}. However, these works are mostly conceptual ideas without concrete frameworks. The ideas of vehicular edge computing (with vehicles as the edge servers) can still be expanded significantly in the future. In the following, we first introduce the computing capabilities of vehicles. Then, we will articulate the computing provisioning in VaaS from two aspects, i.e., computing servers on the move and temporary stationary computing servers. Our design philosophy is in the similar spirit to the emerging computing first networking (CFN), which attempts to leverage computing and communications information to determine a subset of edge servers from multiple geographically distributed edge sites to best serve task computing requests~\cite{krol2019compute}, except that the edge servers considered by us are (mobile) SCCSI-empowered vehicles.

\subsection{Computing Capabilities of SCCSI-empowered Vehicles}
To meet the soaring demands from autonomous driving and in-car infotainment, connected and autonomous vehicles are becoming ``supercomputers on wheels''~\cite{ding2018smart}. For example, NVIDIA DRIVE AGX Pegasus uses the power of 2 Xavier Systems-on-a-Chip and 2 Turing GPUs to achieve 320 TOPS of computing capability, which is built for Level 4 and Level 5 autonomous driving and robotaxis~\cite{NVIDIA}. NVIDIA DRIVE Thor unifies intelligent functions, including automated and assisted driving, parking, driver and occupant monitoring, digital instrument cluster, in-vehicle infotainment, into a single architecture, which will be available for automakers’ 2025 models and achieve up to 2,000 TOPS of performance~\cite{NVIDIAThor}.

In addition to computing for driving operations, auto manufacturers are interested in deploying advanced in-car gaming systems to enhance drivers/passengers' experience, which is particularly essential in the self-driving era when drivers are freed from tedious driving tasks. In response to this need, Telsa has already equipped their cars with Tesla Arcade processor, which is a 10-teraflop gaming system comparable to the latest-generation gaming consoles, to support in-car gaming experience. According to a survey, among drivers younger than 30, 77\% are interested in owning vehicles equipped with VR technology~\cite{Accenture}, which can allow in-car gaming. The great customer demand is expected to further drive auto companies to develop more advanced in-car infotainment systems, thereby endowing the vehicles with more significant computing capabilities for entertainment purposes. All these development trends endow vehicles with powerful computing capabilities that could be harvested.

\subsection{Computing Servers on the Move}
\begin{figure*}
\centering
\includegraphics[width=0.6\textwidth]{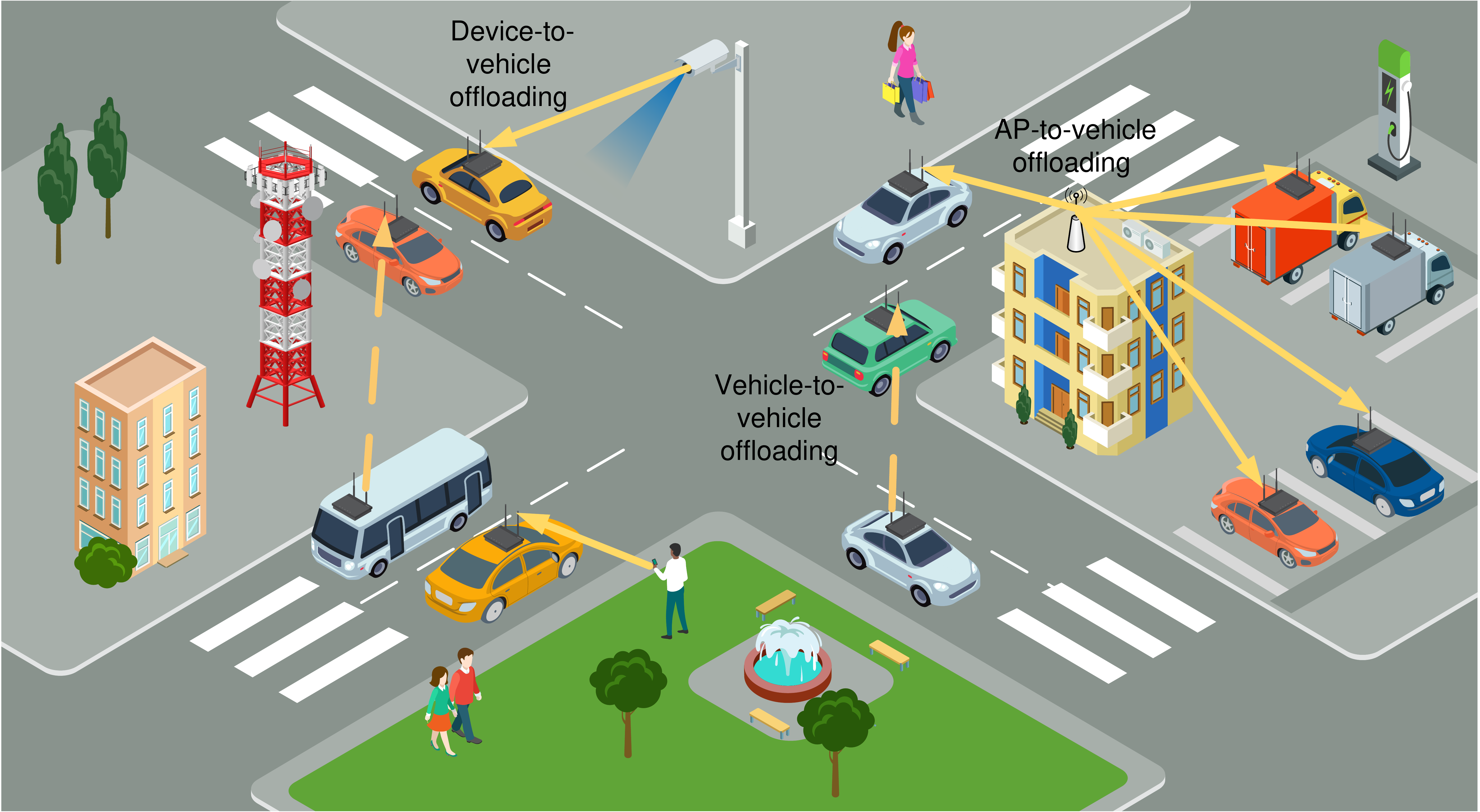}
    \caption{Computation offloading in VaaS, including vehicle-to-vehicle, device-to-vehicle, and AP-to-vehicle computation offloading.  
    }\label{fig:computing}
\vspace*{-0.1in}
\end{figure*}

Vehicles on the move can provide computing services to surrounding end users, including in-car occupants, pedestrians on city streets, and IoT devices, by leveraging their underutilized on-board computing resources. Clearly, this vision is based on the premise that exploiting vehicular computing resources to help others would not disrupt autonomous driving and cause dire consequences. This can be justified since vehicular applications would not run all the time even when they are on the move. For example, entertainment applications may not be on. Also, some optional or advanced driving services, such as navigation, may be temporarily turned off. In these cases, SSP can harvest these resources to provide services without causing safety concerns. When the applications on these vehicles start running, the harvested resources can be returned to the vehicles. In what follows, we will discuss two possible use cases where moving vehicles serve as edge computing servers. For illustrative purposes, computation offloading under various situations is depicted in Fig. \ref{fig:computing}.

\subsubsection{Vehicle-to-Vehicle Offloading} Vehicles can help each other on the road. Vehicles in a vicinity can form a computing cluster to collaborate with other cluster members over V2V links for computing. For instance, when a vehicular user plays a high-end game, this vehicle could leverage the computing resources from the surrounding vehicles to enhance the gaming experience~\cite{palazzi2007online}. This paradigm, called cooperative computing in this paper, generally reckons on reliable and fast data exchange between neighboring vehicles. The high mobility of vehicles, however, may result in frequent service disruption and adversely affect users' experience. To tackle this issue and keep a long enough contact time, the information of moving directions and velocities of vehicles should be taken into consideration. The judicious task offloading and routing decisions can be made based on global network knowledge to ensure a relatively long session between requesting vehicles and serving vehicles. Some papers devise task offloading schemes for vehicular edge computing where one CAV can outsource its tasks to surrounding vehicles. By identifying that there are idle resources on moving vehicles, Feng et al. propose a workflow to support the autonomous organization of vehicular edge nodes and design task assignment policies between neighboring vehicles without requiring roadside infrastructure~\cite{feng2017ave}, where task assignment is done in a distributed manner. A scheduling algorithm based on ant colony optimization is proposed to solve the NP-hard job-assignment problem with fast convergence. In \cite{zhu2018fog}, Zhu et al. investigate task offloading from client vehicles to server vehicles with the assistance of cellular infrastructure, where a base station serves as the coordinator and relay to facilitate task delivery. A task allocation problem is formulated to minimize the service latency and quality loss (i.e., resolution degradation in video tasks). In \cite{su2018distributed}, Su et al. develop a market-based optimal computing resource allocation problem that allows server vehicles to sell their computing power, where a SDN architecture and a star topology are considered in which a centered vehicle outsources its computing tasks to the surrounding vehicles for autonomous driving. While these works make a good attempt to enable vehicle-to-vehicle task offloading, this area still demands further exploration. For example, most of the works along this line rely on simplified network models without considering spectrum allocation among vehicles~\cite{deng2022actions}. Furthermore, multi-hop task delivery and routing among vehicular networks, which is crucial for better load balancing, have not been well studied as yet~\cite{deng2023task,deng2021how}. 

Although the aforementioned works tackle general V2V task offloading, computing offloading in highly dynamic vehicular environments is always challenging. One special case on the road that can easily achieve cooperative computing is platooning. Since cars within a platoon could drive themselves a meter apart, platooning leads to fuel economy due to reduced air resistance and enhanced traffic efficiency~\cite{jia2015survey}. In fact, platooning also yields significantly shortened and nearly constant distances between vehicles, thereby boosting the reliability and throughput for V2V communications while minimizing energy consumption~\cite{sridhar2021piecewise}. For this reason, high-speed and reliable communications can be achieved within a vehicular platoon with line-of-sight (LoS) channels, thereby enabling effective cooperative computing among a group of vehicles.

\subsubsection{Device-to-Vehicle Offloading} 
Moving vehicles with SCCSI capabilities can also provide edge computing services to roadside mobile users or IoT devices when passing by. Given the fact that vehicles are moving but roadside devices are generally immobile or slowly moving, vehicular mobility poses great challenges to device-to-vehicle offloading due to the short contact time between requesting devices and mobile servers, which calls for the new design of computation offloading.

By taking advantage of vehicular mobility, some research efforts have been made to offload computing tasks from roadside users to mobile vehicles. In \cite{sun2019adaptive}, Sun et al. propose an adaptive learning based task offloading algorithm based on the multi-armed bandit theory in order to minimize the average offloading delay. Three task offloading modes are considered therein: vehicle-to-vehicle offloading, pedestrian-to-vehicle offloading, and vehicle-to-infrastructure offloading. In \cite{pang2019spath}, we propose a framework to find the safest walking path in a city by offloading video frames from street cameras to selected vehicles to perform surveillance data analytics. A joint optimization problem of computing resource allocation and computing task assignment is solved to minimize the end-to-end service latency. In \cite{chen2022timeliness}, we envision that roadside IoT devices, such as wireless cameras, can offload computing tasks to vehicles passing by, and devise timeliness-aware reverse auction mechanism design to select and incentivize vehicles to maximize the system utility within a given budget by comprehensively considering vehicles' future trajectories, travel time, and ask prices. 

One special case on the road suitable for device-to-vehicle offloading is when vehicles are slowly moving or there is a traffic jam. In such cases, the speed of vehicles can be greatly reduced. While traffic congestion is widely considered as a nightmare, interestingly, it may alleviate another kind of congestion, i.e., network or ``computing" congestion. Thanks to the slow speed, reliable communications can be easily established between roadside devices and vehicles because of their increased contact time. Roadside devices can therefore transmit large chunks of data to a vehicle by taking advantage of this extended contact period. Moreover, since the vehicle density on the road increases in such cases, roadside devices have better opportunities to find available computing servers. In addition, the inter-vehicle distances are significantly shortened such that high-speed communications can be carried out between vehicles, which facilitates task routing and load balancing among vehicles.

The aforementioned approaches employ direct short-range communications between devices and vehicles for task offloading, thereby bypassing the use of base stations and saving cellular bands. However, as we discussed, the short contact time between vehicles and roadside users makes the design quite challenging, especially under high-speed vehicular networks. To alleviate this issue, one promising method is to utilize base stations or roadside units as intermediate relay nodes to set up connections between devices and vehicles~\cite{zhou2019exploiting,zhou2019computation}. Specifically, the roadside devices first transmit task data to the base station, and then the base stations forward the data to proper vehicles with available computing resources. Due to the powerful communication capabilities of base stations, the base-station-assisted scheme extends the communication range and prolong the device-vehicle contact time. This scenario is illustrated as ``access point (AP) to vehicle offloading'' in Fig. \ref{fig:computing}. In \cite{zhou2019exploiting}, various architectures for vehicle fog networks, including vehicle-vehicle offloading, vehicle-RSU-vehicle offloading, and pedestrian-RSU-vehicle offloading, are reviewed. Two case studies are provided. In the first case study, reinforcement learning approaches and coded computing are combined to improve the service reliability for V2V task offloading. In the second case study, for RSU-assisted task offloading, a task replication policy is developed to minimize the deadline violation ratio by assigning one task to multiple vehicles. In \cite{zhou2019computation}, by considering RSU-assisted task offloading, a contract-theory-based incentive mechanism is devised to maximize the expected utility (i.e., reduced service delay) of the base station, where different contract items are offered to vehicles with heterogeneous resource availability. A caveat of these RSU-assisted schemes is that the uplink and downlink access bands of RSUs are employed to relay the tasks from end devices to vehicles, resulting in low spectrum utilization compared with direct vehicle-vehicle offloading.

\subsection{Temporary Stationary Computing Servers}
Compared to moving vehicles, exploiting stationary SCCSI-empowered vehicles can provide more stable edge computing services. Cars are parked 23 hours a day on average~\cite{RAC2021cars}. When these ``supercomputers'' are parked, there would be massive computing resources idle, particularly when they are autonomous vehicles. Harvesting these resources is less challenging than the cases of moving vehicles. On the one hand, the computing pool formed by stationary vehicles reduces the uncertainty in resource availability. On the other hand, energy consumption is also not a big concern due to the power supply. Since the share of electric cars continues growing (for example, Europe is forecast to have 67.3\% electric vehicles by 2030~\cite{Citi2023state}), vehicles can get charged when parked, thereby providing services without draining batteries. In what follows, we elaborate on two scenarios employing stationary vehicles as edge computing servers for service provisioning in smart cities.

\subsubsection{Parked Vehicles as Edge Computing Servers} 
Parked vehicles are ubiquitously available in urban areas, providing plentiful idle computing resources for edge computing. If managed properly, vehicles parked at parking lots can form computing clusters with significantly powerful aggregated computing resources such that city-wide IoT data can be transported there for processing and computing. 

The idea of leveraging parked vehicles as computing servers has attracted intensive attention from academia. As alluded before, the concept of vehicular cloud/fog computing has been suggested a long while ago \cite{abuelela2010taking,eltoweissy2010towards,olariu2011taking,hou2016vehicular}. They discuss the application scenarios for parked vehicles as servers, such as data cloud in a parking lot and data center at a mall. Nevertheless, these works still mostly focus on conceptual elaboration without considering systematic design. In auto industries, harnessing computing resources on parked vehicles has drawn some attention recently. For example, Canadian scooter and automotive maker Daymak is developing software to enable autonomous vehicles to mine cryptocurrency when vehicles are parked for extended time period~\cite{Daymak2021Daymak}. However, these are very initial attempts and there exist many practical design problems, including the integration of communication and computing design. 

Different from traditional cloud/edge computing with dedicated computing servers in place, harvesting parked cars' resources is subject to the resource availability and the presence of vehicles. SSP still needs to deal with the volatility of vehicles in order to deliver reliable services. Particularly, cars in a parking lot come and go, which impacts the resource availability. In \cite{han2018dynamic}, by observing that the arrivals of computing tasks and locations of vehicles are uncertain, we have developed a dynamic pricing strategy to incentivize parked cars to share computing resources to handle computing requests. The number of vehicles in a parking lot is modeled as a stochastic process with arrivals and departures whose rates can be obtained from historical information. In \cite{zhang2019parking}, an auction-based incentive mechanism has been devised for parking reservation to lead on-the-move vehicles to available parking places with less effort while exploiting the computing capability of these parked vehicles for service provisioning. A multi-round multi-item parking reservation auction is proposed to attract the on-the-move vehicles with computing capabilities via monetary rewards. In \cite{peng2020multiattribute}, a double auction based incentive mechanism has been proposed for vehicular edge computing to incentivize parked vehicles to share their computing resources of multiple attributes, such as locality, reputation, and computing power. In \cite{ma2021parking}, parked vehicles are organized in different clusters (virtual edge servers) for task execution. Specifically, a task scheduling algorithm that jointly determines edge server selection and resource allocation has been proposed. Moreover, after assigning tasks to a virtual edge server, a local task scheduling policy is developed to determine which parked vehicles to process the task by taking into account the remaining task delay and the battery energy levels of parked vehicles.

Although these works serve as good attempts to exploit the potential of the computing power of parked vehicles, there still lacks an effective approach to the joint design of communications and computing, as envisioned in our proposed SCCSI service network. In the aforementioned works, the underlying assumption is that bandwidth limitation is not a concern and therefore they only focus on the computing aspect, which nevertheless, is not realistic. To push this vision into reality, there is an urgent need to design integrated networking and computing schemes to establish high-speed networks for computing task delivery between end users and the pool of parked vehicles. Along this line, the proposed SDN-based approach is suitable for such an integrated design as SSP has global network knowledge for task assignment, channel allocation, and data routing by considering network conditions. It is noted that channel allocation and task routing are crucial to the system performance, as we intend to enable concurrent transmissions in the highly dense parked vehicular networks. The detailed schematic design can be on a case-by-case basis. For example, for a lot with long-term parking at the airport far from the major population, computing tasks can be delivered there through backbone networks, after which base stations there download tasks to parked vehicles for distributed computing. In contrast, for cars parked along the roadside, short-distance communications (i.e., device-to-vehicle communications) can be directly carried out between parked vehicles and citizens/roadside IoT devices for task offloading, where task migration and service reliability should also be carefully investigated due to the typically short parking period on the roadside. Furthermore, various task models can also lead to different designs. For instance, unlike the prior works in which one task is processed by one vehicle, a compute-intensive task can be partitioned into multiple pieces so that a pool of parked vehicles can collaboratively perform a task to reduce the latency.

\subsubsection{Stopped Vehicles as Edge Computing Servers} 
Due to urban planning, parked vehicles are generally concentrated in certain regions. To extend computing service coverage, we could consider another special case in signalized intersections where there always exist stopped vehicles in the direction(s) at the red lights. The group of vehicles stopped at a signalized intersection naturally form a temporarily stationary cluster of PoCs. In our previous work, we have exploited this phenomenon to assist data transportation and caching in VANETs~\cite{ding2018virtual}. It would be interesting to investigate how this cluster of the temporarily stopped SCCSI-empowered vehicles could provide other typed SCCSI services at intersections. 

In smart cities, we envisage that stopped vehicles at intersections will play a vital role in computing. First, there will be huge computing demands at intersections. According to the U.S. Department of Transportation, roughly half of all injury crashes and one quarter of all fatal crashes in the United States occur at intersections~\cite{Intersection}. To make intersections safer, sensor information sharing and cooperative computing are critical. By harnessing the idle resources, the stopped vehicles can perform computing by aggregating and analyzing the surrounding vehicles' captured sensor data to effectively reduce the visual occlusion and extend everyone's awareness beyond their field of view at intersections. The cameras/sensors deployed at intersections can also transmit their data to stopped vehicles for processing to realize traffic monitoring and safety surveillance. Second, stopped vehicles generally have plentiful idle resources, particularly for the autonomous vehicles, which can be utilized for service provisioning. Before the traffic light turns green, it is safe for vehicles to pause their most autonomous driving operations, leaving computing resources for other use. At last, compared with parked vehicles, stopped vehicles are in close proximity to many roadside devices, thereby lowering service latency and reducing communication cost. While parked vehicles constitute the majority of stationary vehicles, many of them are located at private garages or underground parking lots perhaps away from the road, and thus reaching these computing sites may incur additional bandwidth costs and extra delays. In view of the aforementioned advantages, the idea of employing stopped vehicles at intersections to provide computing services has been mentioned in \cite{sun2020novel}, where Sun et al. propose to optimize signal control at traffic intersections for creating vehicular cloud computing by maintaining the number of vehicles at intersections always beyond a desired threshold. Nevertheless, the discussion does not involve system design and optimization of communications and computing.

Of course, there are still many research problems open. The first problem is how to enable effective task migration to and from vehicles and hence redistribute tasks when vehicles arrive and depart. When the vehicles processing tasks start to move, they can migrate the unfinished tasks to the vehicles stopped in the other direction(s) to guarantee service continuity. In this case, the model needs to consider the migration burden, especially the communication cost. The second problem is how to effectively utilize the dynamic  computing resource availability at intersections. During the red traffic light, vehicles will continue arriving, increasing the computing power in the computing pool. Thus, the design for computation offloading should take the incoming vehicles into account. When there is insufficient computing power on stopped vehicles to handle computing requests, the computing tasks can be cached at these stopped vehicles first and then forwarded to incoming vehicles after they arrive and stop. 

\section{Storage as a Service\label{storage}}
To enable fast data retrieval, edge caching has attracted significant attention~\cite{li2018survey}, referring to storage at the edge of networks in order to provide low-latency data delivery to consumers. For example, popular contents, such as trending videos in a social platform, can be stored at the network edge before users' requests, enabling fast downloading once they are requested. Also, different kinds of AI models can be stored at the edge and pushed to users for performing computations once they request. However, unlike cloud storage typically assumed to have significant storage space for storing the entire data library, edge caching features limited storage. VaaS can relieve this issue because SCCSI-empowered vehicles can serve as cache helpers to considerably increase the data storage capacity at the network edge. Note that in this section, we focus on the idea of leveraging vehicles as cache helpers, which is different from the study on content caching for vehicular ad hoc networks, such as \cite{su2018edge}, where RSUs cache data for vehicular users.

\subsection{Parked Vehicles as Edge Caching Nodes}
\begin{figure*}
\centering
\includegraphics[width=0.6\textwidth]{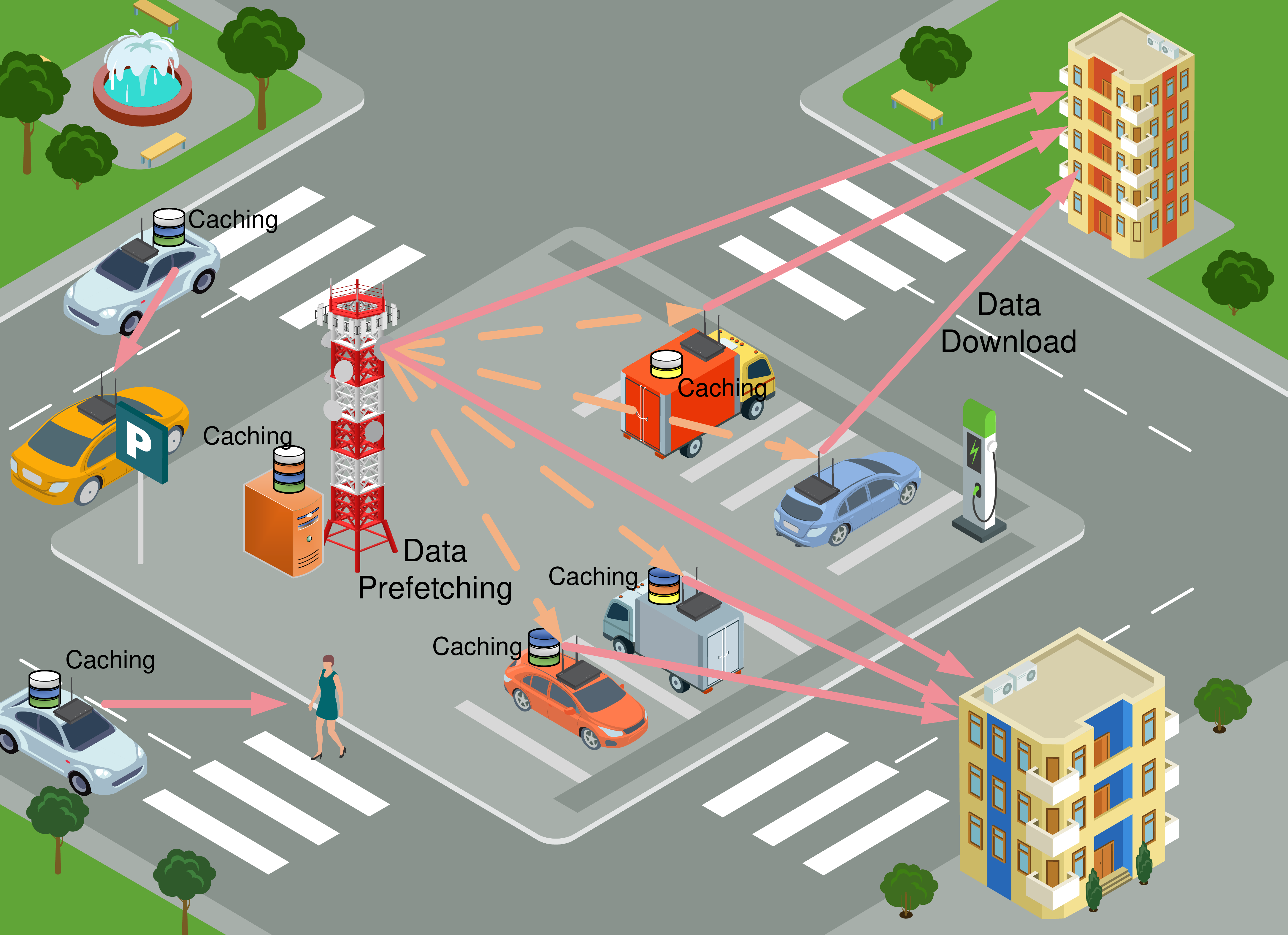}
    \caption{Vehicle-empowered edge caching in smart cities, where 
    both parked and moving vehicles serve as cache helpers to directly download data to nearby users upon request.  }\label{fig:edgecaching}
\vspace*{-0.1in}
\end{figure*}
Parked vehicles can form temporary storage infrastructure due to their large data storage space and idle communication capabilities. Once a user submits the same content request, vehicles with the content in place can directly deliver it to a user without needing to retrieve the content from remote servers and even base stations, as depicted in Fig. \ref{fig:edgecaching}, substantially relieving the burden on radio access and core networks. 

As in traditional edge caching, there can also be two ways for vehicles to cache these contents, i.e., reactive caching and proactive caching, regarding whether to cache a content after or before it is requested~\cite{yao2019mobile,zhang2017cooperative}. For reactive caching, vehicles store popular contents because they have served as relay nodes or data consumers for these messages. Also, when base stations push contents to end users in a multicasting/broadcasting mode, vehicles parked nearby can also overhear and cache the contents. For proactive caching, base stations push the contents to vehicles before users actually demand. Reactive caching reduces backhaul costs as proactive caching might fetch contents that will never be consumed. However, reactive caching works poorly for cases where fewer users request the same content. The selection of caching mechanisms is dependent on many factors, such as backhaul capabilities and spatio-temporal content/service popularity. In \cite{elsayed2022predictive}, Elsayed et al. have proposed a predictive proactive caching scheme to pre-cache the data at parked vehicles before user arrives by exploiting the daily driving routine and predictable behavior of vehicular users. A travel time prediction scheme that incorporates the use of a Long Short-Term Memory (LSTM) network, trained using particle swarm optimization, is developed for travel time estimation, based on which a content placement scheme is proposed to maximize cache hits by considering the mobility of content requests. To compensate for the resource consumption for cache helpers, in \cite{su2016game}, by assuming that some parked vehicles store requested contents, Su et al. have developed an incentive mechanism to maximize the utilities of moving vehicles (content requesters), RSUs, and parked vehicles (content providers). The interactions between RSUs, parked vehicles, and moving vehicles, is modeled as a Stackelberg game, where the former two serve as the game leaders to determine the prices while the latter chooses the percentage of content requested from each side according to the claimed prices.

In-network cached data should be managed and distributed to users in an appropriate manner, which is a non-trivial task. It is widely recognized that information-centric networking (ICN) can speed up the retrieval of contents~\cite{dabirmoghaddam2014understanding,yue2014dataclouds}. Different from the traditional IP-address-based networking, ICN is a content-centric paradigm where users can acquire content using the content's name without explicitly referring to the IP of a source host. In VaaS, the popular spatio-temporal contents, such as trending videos, advertisements, and road traffic information, or computing-related data, such as AI models and data libraries, are widely distributed over networks so that consumers can fetch the needed content from proximate vehicles via one-hop or multi-hop communications. Given the dynamics of parked vehicles, ICN offers additional benefits by effectively coping with the arrivals and departures of vehicles. Specifically, benefiting from pervasive in-network caching, a user can obtain the desired content from multiple resources based on anycasting, which is robust to dynamic network environments. 

ICN mainly works as a distributed and best-effort framework, implying that there may lack QoS guarantee for content delivery. To satisfy the stringent QoS requirements in 5G+, the combination of ICN and a certain level of centralized network control is beneficial. In the SCCSI service network, a fixed PoC (e.g., at a base station or an AP) can easily gather content distribution information from parked vehicles within its coverage, because it serves as the gateway for pushing the data to vehicles. Therefore, we can enable a base station to store the bindings from object names to the cache helper's identities by maintaining a lookup table. Once the content is found in one of the cache helpers (i.e., parked vehicles), the fixed PoC can act as a central controller to make routing decisions and configure forwarding rules based on the aggregated information around the network to ensure the end-to-end QoS for content retrieval to the user. If the requested content is not found in its lookup table, the fixed PoC then forwards the requests to nearby PoCs, following the idea of ICN. 

\subsection{Moving Vehicles as Edge Caching Nodes}
Moving vehicles can also serve as cache helpers. There are several promising scenarios benefiting from employing moving vehicles as cache helpers. First, vehicle occupants may repeatedly consume popular or location-aware contents, including trending multimedia contents and HD maps, which could directly be fetched from adjacent vehicles. Second, public transits, including buses and metro subways (metros for short), can store popular contents, such as daily news and TV series, for delivery to passengers upon requests. Given a large number of passengers, the backhaul links from buses/metros to base stations are typically the communication bottlenecks. Thereby, caching the trending contents in advance could dramatically speed up content downloading for users on public transits and reduce the traffic over radio access networks. 

Given high mobility of cache helpers in the aforementioned case, fully centralized management (e.g., a base station maintaining the full content list at cache helpers) might be too costly due to highly frequent network changes. For this reason, ICN-enabled distributed caching arguably provides the most practical solution where users (e.g., occupants in vehicles) send out content requests to adjacent moving vehicles to check whether contents are around. Along this line of thoughts,  Vigneri et al. have devised a caching policy based on a model where a user sends a content request to nearby vehicles, and if not found, the request is redirected to the cellular infrastructure~\cite{vigneri2017quality}. In \cite{zhang2019mobility}, by employing ICN, Zhang et al. have developed an online vehicular caching design to optimize network energy efficiency. The interaction between caching vehicles and mobile users is modeled as a two-dimensional Markov process to characterize the network availability of mobile users. The caching decision-making process is formulated as a fractional optimization model, aiming to achieve optimal energy efficiency. By employing nonlinear fractional programming techniques and Lyapunov optimization theory, they derive a theoretical solution for this model and devise an online caching algorithm to facilitate the implementation of optimal vehicular caching. In \cite{wang2021information}, Wang et al. have proposed a proactive in-network caching scheme for heterogeneous network nodes, i.e., vehicles and RSUs, where vehicular users can request contents from both of them based on ICN. Based on this model, they formulate a QoE optimization problem to minimize waiting time for content retrieval under cache capacity limitations. The optimization problem is decoupled into two sub-problems, namely, the power allocation problem and the content placement problem, which enables the acquisition of the optimal QoE.

\section{Intelligence as a Service\label{intelligence}}
\begin{figure*}
\centering
\includegraphics[width=0.6\textwidth]{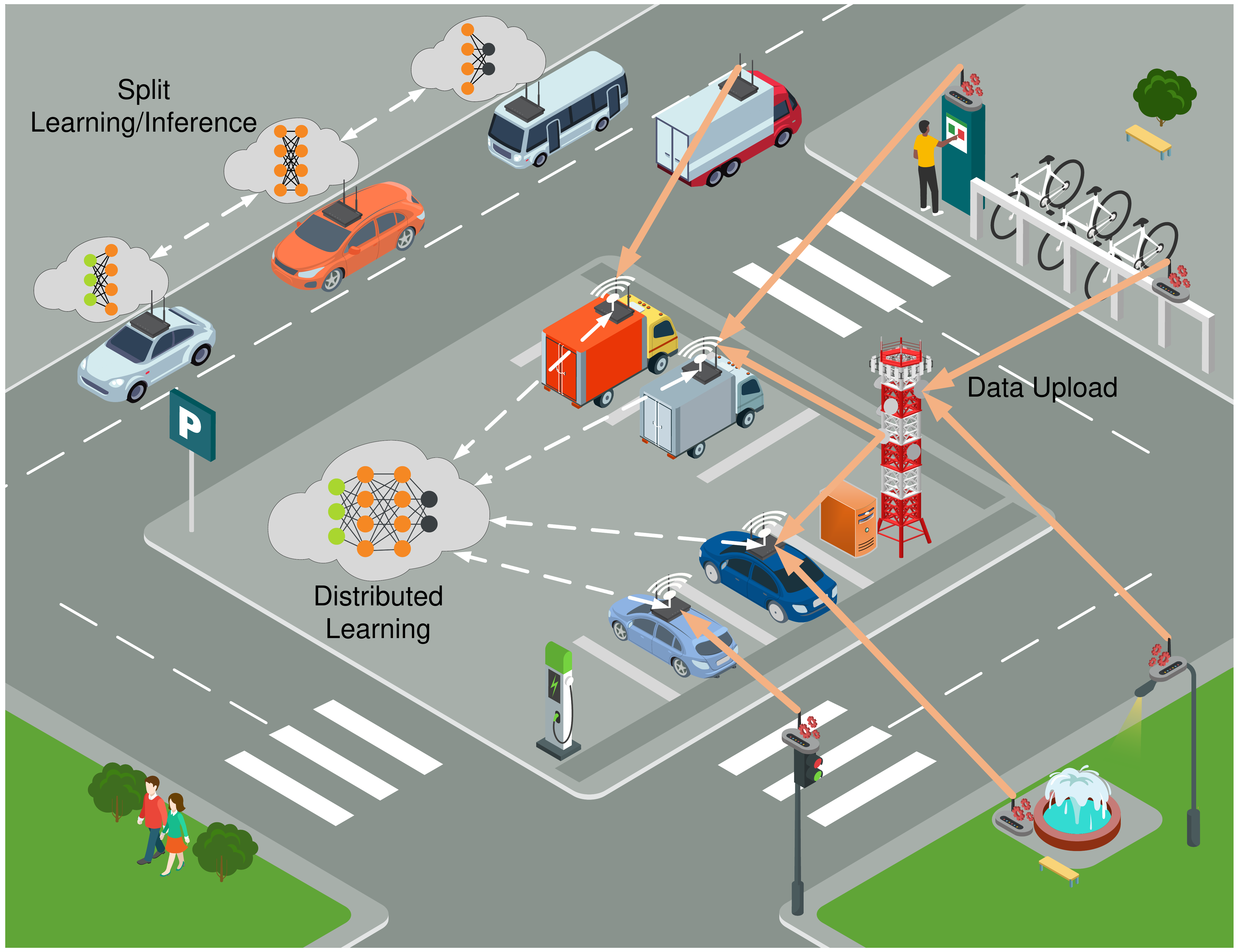}
    \caption{Vehicle-empowered edge intelligence in smart cities.
    In a parking lot, vehicles perform distributed learning based on their own data or data collected from smart city environments. In a  vehicular platoon, multiple vehicles conduct collaborative learning/inference based on split machine learning.}\label{fig:edgeintelligence}
\vspace*{-0.1in}
\end{figure*}

AI-empowered functionalities pervade almost all contemporary electronic products, including vehicles and other IoT devices in smart cities. For example, the current Tesla Autopilot involves 48 neural networks to perform various tasks on cars, such as semantic segmentation, object detection, and depth estimation~\cite{tesla}. However, AI is notoriously data-hungry and compute-intensive, posing significant challenges to the wireless edge. Fortunately, by leveraging SCCSI-empowered vehicles as edge nodes, vehicles can train powerful machine learning models and perform model inference, either for vehicles themselves or pervasive IoT devices in a smart city. In this section, we review vehicle-enabled edge intelligence by elaborating on two aspects, i.e., edge learning (training) and edge inference, as illustrated in Fig. \ref{fig:edgeintelligence}. We will mainly focus on how to crowdsource vehicles as mobile edge servers to provide services in smart cities. However, in-vehicle ML, where a vehicle performs ML tasks (e.g., self-driving) for itself, will not be discussed here due to page limitation although it is important part of edge intelligence. 


\subsection{Vehicle-Enabled Edge Learning}
Edge learning refers to AI model training at the network edge. Massive data will be generated in smart city environments, such as surveillance videos, traffic information, environmental sensory data, and utility data, which can be delivered to and trained by the pervasive vehicles to continuously extract intelligence and improve smart city operations. However, since model training is generally compute-intensive, latency-sensitive model training requires load balancing among multiple vehicles (i.e., dynamic edge nodes). Unlike fixed cloud/edge servers where multiple servers are usually interconnected via high-speed fiber links, the computing resources on vehicles are naturally distributed and can only be interconnected via wireless communications. This brings design challenges and also opens research opportunities in how to pool the distributed vehicular resources together for edge learning. In what follows, we classify vehicle-enabled edge learning into two categories, i.e., data-split learning and model-partition learning, according to how multiple vehicles collaborate.

\subsubsection{Data-Partition Learning} To harvest vehicles' computing resources, we could employ data-partition learning, where multiple vehicles collectively train the same AI model with different sets of data samples. To achieve this, there are two approaches. In the first approach, e.g., federated learning (FL)~\cite{xu2014control,xu2015privacy,xu2017my,mcmahan2017communication} or its variants, an application server pushes an initial model to vehicles, where the vehicles leverage their private data and computing resources for training, as suggested in \cite{posner2021federated,du2020federated}. The locally updated models will be uploaded and aggregated to obtain an improved global model. The process will repeat until the model converges. FL aims to train a global model while ensuring that private data never leaves local devices (e.g., vehicles) to enhance their privacy. The second approach, i.e., the traditional distributed learning, is similar to federated learning, except that the application server is the owner of training samples and the goal is merely to harvest computing capabilities of vehicles. In this case, the application server can push both the initial model and partitioned data samples to vehicles in order to leverage vehicles' powerful onboard capabilities for collaborative training.

\begin{table*}[htbp]\label{table3}
\centering
\caption{Related works on distributed machine learning over vehicles. In these works, vehicles collaboratively train AI models or perform model inference by utilizing their distributed resources.}
  \renewcommand{\arraystretch}{1.4}{
  \setlength{\tabcolsep}{2mm}{
\begin{tabular}{|c|c|c|c|}
\hline
\textbf{References}                                         & \textbf{Type}&\begin{tabular}[c]{@{}c@{}}\textbf{Description}\end{tabular} & \textbf{Highlights}
\\ 
\hline
\cite{xiao2021vehicle}& Federated learning & \begin{tabular}[c]{@{}c@{}}Develop an algorithm to minimize the training delay and energy consumption\\ in vehicle-based FL by jointly optimizing the onboard computing capability,\\ transmit power, and local training accuracy.\end{tabular} & \begin{tabular}[c]{@{}c@{}}Training latency\end{tabular} \\ \hline
\cite{pervej2023resource}& Federated learning & \begin{tabular}[c]{@{}c@{}}Address a joint learning and radio resource allocation problem under delay,\\ energy and cost constraints to maximize the probability of successful reception\\ of local model updates from vehicles.\end{tabular}  & \begin{tabular}[c]{@{}c@{}}Training accuracy\end{tabular}
\\ \hline
\cite{zheng2023autofed}& Federated learning & \begin{tabular}[c]{@{}c@{}}Present a heterogeneity-aware federated learning framework to fully exploit\\ multimodal sensory data on autonomous vehicles to enable robust\\ autonomous driving.\end{tabular} & Data hetergeneity \\ \hline
\cite{zeng2022federated}& Federated learning & \begin{tabular}[c]{@{}c@{}} Design a contract-theoretic incentive mechanism to speed up the model\\ convergence by considering the time-varying participation of vehicles.\end{tabular}  & Incentive design\\ \hline
\cite{pokhrel2020federated}& Federated learning & \begin{tabular}[c]{@{}c@{}} Propose blockchain-based federated learning for autonomous driving to enable\\ decentralized on-vehicle training.\end{tabular} & Decentralization\\ \hline
\cite{jouhari2021distributed}& (Layer-wise) Split inference & \begin{tabular}[c]{@{}c@{}} Develop collaborative split inference among a group of UAVs as with the\\ aim of minimizing the classification latency under limited resources.\end{tabular} & Inference latency\\ 
\hline
\cite{liu2023accelerating}& (Horizontal) Split inference & \begin{tabular}[c]{@{}c@{}} Employ horizontal model partitioning to accelerate deep neural network inference\\ with reliability guarantee by considering vehicular mobility.\end{tabular} & Inference reliability\\ \hline
\end{tabular}}}
\end{table*}

Implementing federated learning over vehicular networks is quite challenging due to the high mobility and short sojourn times of vehicles, imposing signficant challenges to model upload in federated learning. Without appropriate vehicle selection and radio resource allocation, roadside units cannot successfully receive the model updates from vehicles. To tackle these challenges, there are a handful of works that optimizes federated learning for vehicular networks. In \cite{xiao2021vehicle}, Xiao et al. have developed an algorithm to minimize the training delay and energy consumption in vehicle-based FL by jointly optimizing the onboard computing capability, transmit power, and local training accuracy. They incorporate the vehicle's position and velocity into the decision-making process to ensure that each vehicle can complete at least one round of training and update it while remaining within the coverage area of the current edge server. They also propose a greedy algorithm that prioritizes the selection of vehicles with superior image quality, which minimizes the overall cost of FL. In \cite{pervej2023resource}, observing that model convergence benefits from the successful reception of models, Pervej et al. address a joint learning and radio resource allocation problem under delay, energy and cost constraints to maximize the probability of successful reception of local model updates by taking vehicles' dataset sizes and sojourn periods into consideration. In \cite{zheng2023autofed}, considering the practical multimodal sensory data from vehicles, Zheng et al. present a heterogeneity-aware federated learning framework to fully exploit multimodal sensory data on autonomous vehicles to enable robust autonomous driving. An autoencoder-based data imputation method is proposed to fill in missing data modality (of certain autonomous vehicles) with the available ones. To further reconcile the heterogeneity, the authors develop a client selection mechanism by exploiting the similarities among client models to improve both training stability and convergence rate. Unlike the aforementioned schemes assuming that vehicles voluntarily contribute to the training process, in \cite{zeng2022federated}, Zeng et al. have designed a contract-theoretic incentive mechanism to speed up the model convergence by considering the time-varying participation of vehicles. In particular, the impacts of varying CAV participation in the federated learning process and diverse data quality of vehicles on model convergence are thoroughly examined. Utilizing this analysis, an incentive mechanism based on contract theory is designed to enhance the convergence speed of federated learning under vehicular environments. To encourage vehicles' participation and eliminate the need for a global server, in \cite{pokhrel2020federated}, Pokhrel et al. propose blockchain-based federated learning for autonomous driving to enable decentralized on-vehicle training. This framework offer rewards proportional to the size of local data samples offered to vehicles and facilitates efficient communication among autonomous vehicles, where local learning modules exchange and verify their model updates in a fully decentralized fashion by exploiting the consensus mechanism of blockchain. A comprehensive analysis of the end-to-end latency is also conducted, which provides insights for deriving the optimal block arrival rate.

In summary, FL design for vehicular networks considers several unique challenges. For example, vehicle selection must be smartly done considering the prediction of their future trajectories. Once a vehicle moves far away, it incurs higher transmission costs and delays to transfer the trained model back to the aggregator. Additionally, wireless channels in vehicular environments are more unstable than stationary cases due to their mobility and blockages (e.g., trucks), FL design may have to take link reliability into consideration to investigate how the transmission failure due to the high-mobility environments will impact model training. 

To fully exploit the potential of data-partitioning training, more aspects are worth exploring. First of all, if data is centralized (for the second approach above), SSP can first shuffle the data samples and then judiciously push the data to vehicles. Data shuffling can make the datasets on vehicle workers be independent and identically distributed (IID) as much as possible, otherwise, a highly non-IID data distribution can severely hinder the model convergence as discovered in federated learning~\cite{li2019convergence}. The dataset should also be partitioned into appropriate sizes for assignment to different vehicles by considering their communications and computing capabilities. Second, SSP must appropriately determine the training epochs (the number of local training iterations before global aggregation) to achieve the tradeoff between model convergence rate and transmission cost. In general, a smaller number of epochs implies more frequent model exchange yet facilitates model convergence. In a spectrum-congested vehicular network, the SSP can increase the training epoch time, asking vehicles to ``train'' more and ``talk less''. At last, multi-hop FL model aggregation and routing can be explored for vehicular networks to reduce service coverage and shrink data traffic, analogous to our previous work for static multi-hop 
FL networks~\cite{chen2022federated}. 

\subsubsection{Model-Partition Training} The other way to realize collaborative learning is model partitioning, where multiple edge servers (e.g., vehicles) share successive sub-models to collectively train a global model based on split learning (SL)~\cite{vepakomma2018split,thapa2022splitfed}. The integration of split learning and wireless networks is elaborated in \cite{lin2023split}. In vehicular networks, we can employ SL in a vehicular platoon, as depicted in Fig. \ref{fig:edgeintelligence}. To execute forward-propagation and back-propagation, two consecutive vehicles exchange the smashed data at the cut layer. SL is a judicious way to leverage multiple edge servers to collaboratively train a global model because it does not need to periodically synchronize models among edge servers (vehicles) as in other distributed learning technologies. Instead, it only requires smashed data exchange between successive nodes, which is typically much smaller than models (e.g., the average layer output size for VGG16 is around 0.1 MB, while the model size is 528 MB). However, it is noted that the total amount of smashed data exchange increases with the size of the training data size. Thus, model-partition training is preferable if the training dataset is relatively small while the model is large. 

In SL, split-layer and sub-model placement greatly impact training latency. Specifically, split-layer decisions result in not only different training workloads partitioned between end devices and edge servers but also varying communication overhead due to the different output data sizes across layers~\cite{lin2023efficient}. A few works have investigated model split for SL over wireless edge networks~\cite{lin2023efficient,wu2023split}. Multi-hop split inference/learning for a given network topology has also been studied with the objective of latency minimization~\cite{tian2022jmsnas,wang2021hivemind}. Given the massive number of vehicles on the road, it is worthwhile to address the joint multi-hop layer split and model placement problem over a mesh of vehicles to minimize the end-to-end training latency (comprising communication and computing delays) for better leveraging of vehicles' resources. Besides, since vehicles can join and leave SL (e.g., vehicles joining and leaving a platoon), model migration would be an interesting topic to migrate the models from a leaving vehicle to other vehicles to continue training, similar to service migration in edge computing~\cite{wang2019dynamic}.



\subsection{Vehicle-Enabled Edge Inference}
After edge training, well-trained AI models can be deployed on edge servers for inference (i.e., prediction). In the context of vehicle-enabled edge inference, input data from end devices are delivered to surrounding vehicles with the well-trained AI model for calculating the outputs (e.g., prediction results). This scenario is a special case of a general vehicular edge computing paradigm in Section \ref{computing}, except the adopted programs are AI models. Many smart-city applications require multi-vehicle collaborative perception to achieve enhanced coverage and robustness. For example, in autonomous driving, multi-vehicle perception can extend the sensing range to ensure safer and smoother driving. In public safety applications, multi-vehicle perception can better detect a crime and identify the face of a suspect. 

A single vehicle may not possess sufficient computing resources as powerful as a fixed PoC. To accelerate edge inference, split edge inference \cite{teerapittayanon2017distributed} can balance the computing workload among multiple vehicles. Multi-hop split inference bears similarity to multi-hop split learning (training) in many aspects except that it does not contain a back-propagation process. In \cite{jouhari2021distributed}, by considering layer-wise model splitting, Jouhari et al. formulate collaborative inference among a group of UAVs as a non-linear optimization problem that aims to minimize the classification latency under limited resources. They consider both homogeneous and non-homogeneous mobility models, referring to the scenarios where the relative distance between UAVs remains unchanged or not. For the non-homogeneous case, based on a mobility prediction method, an optimization problem is solved by taking into account the future locations of UAVs. Apart from layer-wise model partitioning, another way is to horizontally divide the model into sub-models to accelerate inference. This method overcomes the limitation of layer-wise splitting which is executed in a serial manner, thereby speeding up inference via parallel computing. Along this line, in \cite{liu2023accelerating}, Liu et al. employ model partitioning to accelerate deep neural network inference with reliability guarantee by considering vehicular mobility, where a vehicle partitions and offloads its data to surrounding vehicles for processing. By formulating a cooperative partitioning and computation offloading problem, two algorithms, namely the Submodular Approximation Allocation algorithm (SA3) and Feed Me the Rest algorithm (FMtR), are developed for computation allocation with a lower bound guarantee for reliability (offloading success ratio). 

There are still numerous open problems in vehicle-enabled edge inference. An interesting scenario is to implement split inference for platoon-based autonomous driving. Since the most critical view is probably what the head vehicle observes in a platoon, the head vehicle could share the computing load with others in the fleet based on split inference, thereby decreasing inference delay and energy consumption. In general, model inference (e.g., object detection) is more delay-sensitive than model training (e.g., improving a model for object detection). Split inference may violate the deadline constraints if intermediate network links are unstable. Based on split inference, we can endow early exit at each intermediate vehicle to achieve the optimal latency-accuracy tradeoff, as proposed in \cite{teerapittayanon2017distributed}. In general, higher inference accuracy can be achieved by traversing more neural network layers on more vehicles at the cost of longer latency, whereas early exits allow the output at intermediate points of neural networks, thereby reducing the inference time. Still using the example of platooning, vehicles in a platoon can be grouped together to execute multi-hop split inference, where fast yet inaccurate outputs from the front vehicles can be used for prediction/decision-making while slow yet accurate outputs from back vehicles can be employed to correct the predictions/decisions made before. This implementation, allowing vehicles to perform edge inference collaboratively with different levels of delay and accuracy provisioning, can also be applied to other mission-critical applications (e.g., vehicle-enabled surveillance analytics as illustrated in Fig. \ref{fig:publicsafety}) in smart cities.
 
\section{Further Research Problems\label{future}}
Pushing VaaS into reality requires systematic design, including architectural network design and customization, network planning, intelligent resource management and mobility management, incentive mechanism design, security and privacy, and blockchain design. We have already articulated some of the design issues when we describe our VaaS. In this section, we discuss several further open problems and challenges in VaaS.

\subsection{Service Capacity Analysis}
Since ubiquitous vehicles can significantly reduce the deployment cost of 5G/6G infrastructure, one crucial problem is how many infrastructure nodes (e.g., PoCs placed along street facilities such as 5G+ BSs, WiFi APs, and RSUs) and how much CAPEX are still needed given the mobility, density, and capabilities of vehicles on the road. This question can be answered by service capacity analysis.

The achievable network capacity is one of the most important metrics in wireless networks~\cite{gupta2000capacity}. The capacity of vehicular networks has been investigated for various V2V and V2I networks~\cite{ding2018access,chen2017capacity}. Different from conventional vehicular ad hoc networks, the SCCSI service network delivers SCCSI services. How to measure the \textit{service capacity} in terms of end-to-end service provisioning over SCCSI service networks plays an essential role in understanding this architecture and network planning, where the service capacity can be defined as ``the task throughput under QoS constraints''. Once obtaining the service capacity of VaaS, a network operator can deploy fixed infrastructure to support services for residents and visitors. 

Stochastic geometry has been established as a standard tool for modeling and designing wireless networks. Poisson point processes (PPPs) has been extensively adopted for network performance analysis, including MEC systems~\cite{ko2018wireless,hui2022server}. Considering the fact that mobile users and edge sites may not be uniformly distributed, more advanced processes, such as Poisson cluster processes (PCPs), have also been considered for MEC systems to characterize the cluster effects (in the sense that mobile users may be clustered around MEC servers)~\cite{gu2021modeling}. Nonetheless, the prior works on MEC analysis have not considered mobile edge nodes as in our scenario. By employing vehicles as edge servers (nodes), we have to consider the limited contact time between vehicles and end users, and the frequent changes in user association due to vehicles' high mobility. In other words, the analytical service capacity depends on not only the spatial distribution of vehicles confined to road layout but also the contact duration by considering vehicular mobility.

\subsection{PoC Placement for Fixed Partial Infrastructure}
The efficacy of the proposed SCCSI service network largely depends on the location of fixed PoCs and available SCCSI resources. 
As the density of PoCs increases, the capability of the SCCSI service network gets stronger, the end-to-end QoS improves due to shorter distances to the edge, and the efficiency of the harvested SCCSI resources increases due to spectrum reuse and resource sharing. Yet, CAPEX or the deployment cost also increases. Therefore, how to optimally place fixed PoCs to provide sufficient premise capabilities with an economically acceptable cost while making the best use of opportunistic vehicular mobility and multi-dimensional SCCSI resources poses great challenges and should be carefully investigated. 

Node placement problems have been studied in various communication and computing networks~\cite{yue2013spectrum,fan2019cost}. For example, in \cite{yue2013spectrum}, we studied a cost minimization problem for relay placement in cognitive radio networks under spectrum and energy efficiency constraints. In \cite{fan2019cost}, Fan et al. have studied the cost-aware placement of edge computing servers by considering end-to-end service delay. Nevertheless, the full integration of fixed and mobile infrastructure and the nature of multi-dimensional resources (SCCSI) largely differentiate the PoC placement problem in VaaS from existing works. Intuitively, assuming the same service demands, the places with higher vehicle traffic density normally require fewer fixed PoCs to deploy since there are already sufficient resources supplied by vehicles. Given the layout of a city segment with street topology, service demands, and vehicle traffic density, an interesting problem is how to place PoCs to achieve a certain percentage of service coverage at the minimum cost.

\subsection{Mobility Management}
Mobility management is always an essential part of wireless cellular networks. Due to user mobility, seamless handover must be ensured to guarantee service experience when a user moves from one cell to the next. In MEC systems, virtual machines (VMs) or containers may have to be migrated from one edge server to another to enable computing service to be as close as to users on the move. Along this line, Wang et al. formulate service migration as a 1-dimensional Markov decision process to minimize the overall cost by approximating the underlying state space as the distance between the user and service locations~\cite{wang2019dynamic}.  Mobility management for vehicular networks has been comprehensively reviewed in \cite{aljeri2020mobility}.

With multi-dimensional SCCSI resources and the mobility of PoCs, mobility management must take on a new look. Prior works on computing service migration merely consider user mobility~\cite{labriji2021mobility,lu2018fast}. Unlike these scenarios, PoCs in SCCSI-empowered vehicles are also mobile. For this reason, mobility management for VaaS has to incorporate the mobility prediction of both end users and PoCs to achieve a seamless handover. Particularly, when a PoC providing computing services leaves a zone soon, containers or virtual machines can be migrated to next PoC in advance for service continuity within the latency bound. Different from what has been done in 5G+ cellular systems, SCCSI service migration should consider not only the computing workload at the PoCs but also spectrum availability on their routes. Thus, mobility management with multi-dimensional SCCSI resources and vehicular mobility is highly challenging, but vitally important for VaaS.

\subsection{Incentive Mechanism Design for VaaS}
To compensate for the costs of PoCs installed in vehicles, well-designed incentive mechanisms lie in the heart of SCCSI service networks. Traditionally, the maintenance cost of cellular networks comes from various factors, such as resource consumption and logistics. In SCCSI service networks, these costs are shared among vehicle owners. Despite this, the actual expenditure of vehicular owners, such as energy consumption for providing services, is typically unknown to the operators. A vehicle owner can therefore claim a much higher ask price than the true cost, which undermines the cost-effectiveness of SCCSI service networks. To make SCCSI service networks more attractive, we expect that SCCSI service networks not only significantly reduce the infrastructure deployment cost in the 5G+ vision, but also incur an overall operational cost lower than or at least comparable to those of 5G+ networks. This will incentivize industrial stakeholders and city authorities to adopt SCCSI service networks for smart cities.

To maintain a cost-effective SCCSI service network, the first research direction is to release reward/pricing mechanisms resistant to market manipulation. Game theoretic approaches, including contract theory~\cite{zhang2017survey} and auction theory~\cite{chen2022from}, offer effective solutions to elicit the true costs and maximize societal benefit (social welfare) or operator's profit. In contract theory, a service provider offers a contract and then each vehicular user can choose the best contract items to maximize its utility. Auction approaches, on the other hand, enable vehicular users to submit ask prices for service provisioning. In our context, since there could be one buyer (service provider) intending to recruit vehicles and many vehicular owners act as sellers, the reverse auction is a suitable approach to characterizing such a market relationship. Participants can achieve the best outcome for themselves by revealing their true types or costs when the contract/auction mechanism design is truthful or incentive-compatible.

While truthful incentive mechanisms for vehicular crowdsourcing have been investigated in several works\cite{chen2022timeliness,gao2018truthful}, they focus on one aspect, single-dimensional SCCSI provisioning (e.g., vehicular crowdsensing or computing), rather than the incentive design for joint multi-dimensional SCCSI resource provisioning. Unfortunately, in a service network, customers can only be satisfied with an end-to-end service guarantee with all SCCSI resources satisfied. For example, a user demanding a ``video analytics'' service would not be happy if there is computing resource available yet no sufficient transmission spectrum to deliver the video to the server. Thereby, when SSP harvests the needed resources by incentivizing resource owners, it must ensure that the multi-dimensional SCCSI resources can be harvested to meet the QoS requirements (an end-to-end delay requirement) for the served applications. With QoS constraints in place, the incentive problem can be formulated as either a service throughput maximization problem or a cost minimization problem. In \cite{chen2022end,chen2022from}, we proposed a service auction to address the joint problem of network optimization and auction design with both economic and QoS constraints. In this approach, service users bid for ``services'' with end-to-end QoS requirements rather than single-dimension ``resources'' alone. However, due to vehicular mobility, designing a service auction tailored for VaaS is more challenging and requires further research efforts.


The second research direction is to design a point-based system attracting participation in the long run. Existing incentive mechanisms for crowdsourcing systems do not account for the psychological effects at different stages. However, similar to loyalty point-based programs, such as Starbuck membership, flight mileage program, and shopping points in our daily lives~\cite{kim2021emerging}, it is beneficial to consider the psychological effects of vehicle owners to stimulate them to stick to the system and contribute their services for the common good. For example, offering a higher reward to new participating vehicle owners could attract newcomers, while offering an extra bonus or a high-level ``membership'' for vehicles to accomplish a target goal (i.e., fulfilling a required number of tasks) could encourage them to continuously contribute their SCCSI resources. The earned points can be used to purchase city services or even paying utility bills later, thereby improving dynamic spatio-temporal resource sharing among citizens, visitors, and different parties in a city.  An interesting research direction is the systematic design of multi-stage incentive mechanisms by considering citizens' expectations over time.

\subsection{Security \& Privacy}
VaaS leverages SCCSI-empowered vehicles for crowdsourcing. For communications and storage service provisioning, state-of-the-art data encryption can be adapted to ensure data confidentiality and integrity upon data transfer or storage, which is not much different from existing communication systems (e.g., 5G UE relaying)~\cite{lu20205g}.  Nevertheless, vehicle-based edge computing results in significant security \& privacy concerns. A curious vehicular user may inspect the data and deduce sensitive information about the service users. Also, it may not honestly conduct the computation and simply return an invalid result to save computational cost. To address these issues, it is essential to ensure data confidentiality and computation integrity for the success of SCCSI service networks. Specifically, data confidentiality refers to preventing direct user data leakage to servers, while computation integrity refers to the user's ability to verify the correctness of the computing results, which is also called result verifiability. 

To prevent data leakage, one common approach is to require edge servers (e.g., vehicles) to only perform computation on users' encrypted data. Homomorphic encryption (HE) is an essential approach to achieving secure computing, allowing servers to perform computations on encrypted data without decrypting it. Fully homomorphic encryption (FHE) scheme, a general result of secure computation outsourcing for arbitrary computation has been shown viable in \cite{gentry2009fully}. FHE supports an unlimited number of additions and multiplications to be performed directly on encrypted data, thereby enabling arbitrary computations on encrypted data. Moreover, a general mechanism for secure computation outsourcing, which combines fully homomorphic encryption with the evaluation of Yao’s garbled circuits (GCs), achieving both data confidentiality and computation integrity~\cite{gennaro2010non}. However, while the aforementioned works provide general solutions to secure computation outsourcing, these schemes suffer from significant computation burdens. There is a tradeoff between generality and computational efficiency. Another line of research on homomorphic encryption, therefore, focuses on fundamental operations (e.g., matrix multiplication) or specific problems (e.g., linear programming)~\cite{wang2011secure}. For example, somewhat homomorphic encryption can be generally employed to empower model inference/training based on encrypted data~\cite{xu2020secure}. Since these schemes are only limited to addition and multiplication, the nonlinear operations in neural networks, such as sigmoid functions and ReLu functions, are usually approximated as polynomial terms at the cost of sacrificing accuracy. Since the computing capabilities of vehicles are generally less powerful than traditional cloud/edge servers, the key design challenge is how to lower the computing and communication overhead of homomorphic encryption while maintaining satisfactory accuracy.

Besides encryption, another approach is to conduct feature extraction from the raw data and then send the less-sensitive features to edge servers (vehicles) for processing. Split learning and inference, which we discussed earlier, fall into this category. Nevertheless, while this approach naturally enhances privacy without incurring additional computation costs, it is possible for a malicious server to reconstruct the raw data from the activations~\cite{pasquini2021unleashing}. To provide stronger privacy protection, differential privacy~\cite{shokri2015privacy} can be used to add noise to the transmitted features, making it harder to infer sensitive information on the server side. This forms another rich research direction for our SCCSI service networks. 

\subsection{Blockchain for VaaS}
When billions of IoT devices and vehicles are connected, fully centralized architectural design faces several challenges. In particular, VaaS involves a tremendous amount of resource and data sharing and hence service trading. If all the transactions and authentications have to go through centralized server or an intermediate third-party, the process will suffer from long latency, heavy computation, a single point of failure, and less transparency. Yet, enabling SCCSI-empowered vehicles to contribute their resources plays a key role in our VaaS, and thus how to enable potentially untrusted parties to entrust each other to perform service trading in a distributed setting is highly challenging but of paramount importance. Blockchain is the perfect platform to address this issue~\cite{swan2015blockchain}. By employing blockchain, agreements on resource/data sharing and service trading can be performed in a fully decentralized and peer-to-peer manner, and the aforementioned limitations can be overcome. In \cite{jiang2018blockchain}, Jiang et al. employ blockchain technology in vehicular networks to enable decentralized and secure storage of big data. Lin et al. \cite{lin2020blockchain} propose a peer-to-peer computing resource trading system to balance computing resources under dynamic spatio-temporal demands between vehicles. There is also a comprehensive review on the integration of the Internet of Vehicles (IoVs) and blockchain~\cite{mollah2020blockchain}. Unfortunately, most of these works focus on single-dimensional resource trading. How to enable trustworthy multi-dimensional service-oriented service trading over highly distributed blockchain-based SCCSI service networks is very interesting and important, and does demand further research. 

In the future, blockchain-based distributed auction design presents a promising opportunity for the development of VaaS. Service auction, as mentioned earlier, can enable resource sharing and service trading among residents and visitors in VaaS. However, traditional auction schemes with central auctioneers are unscalable. Moreover, a fully trusted ``third party'', as always assumed to serve as the auctioneer, may not exist in many practical application scenarios. To overcome these limitations, distributed auction design with direct peer-to-peer trading can be a potential research direction for VaaS. Double auction, which supports many-to-many trading, will be an appropriate auction approach in our SCCSI service networks where all trading transactions among vehicular resource contributors and service receivers can be recorded over the blockchain. In \cite{thakur2018distributed}, a double auction scheme for peer-to-peer energy trading based on blockchain has been proposed, where a peer can act as an auctioneer while behaving lawfully. 

As a remark, we notice that SCCSI-empowered vehicles most likely will register with SSP or the city authority with their vehicles and profiles in order to join VaaS to participate SCCSI service network activities (contributing resources for rewards or receiving services from SCCSI service networks). Thus, such a set of mobile PoCs (or vehicles), fixed PoCs along the edges of city streets (collocated with RSUs), and backhaul/backbone edge nodes, could form a fundamental trustworthy subnet to help manage the security and privacy. More importantly, SCCSI service networks provide rich communications and computing services, which will be capable of providing the needed communications and computing services for blockchain. This shows another reason why blockchain and our envisioned SCCSI service network are perfect fit. In the future, there is a need to investigate the design of an effective blockchain-based platform that can enable multi-dimensional SCCSI service trading to support smart city operations and services.

\section{Conclusions\label{conclusion}}
In the era of digitalization and intelligentization permissive to all over the places, things, and operations, we anticipate that without exception, vehicles will become more than a way of just a transport, but also the infrastructure to fuel numerous IoT applications to modernize city operations and services. With innovative digital and AI empowerment, vehicles can be leveraged to carry the needed capability to support sensing, communications, computing, storage and intelligence (SCCSI) for city modernization. Therefore, in this paper, we have advocated Vehicles as a Service (VaaS) that leverage vehicles' empowered capabilities to build up such a service network for smart cities. To fully reap the potential of VaaS, we have proposed an SCCSI service network architecture, comprehensively reviewed the aspects of sensing, communications, computing, storage, and intelligence, and identified problems, challenges, and future research directions in such a design. We expect this paper, much more as a position paper, will inspire more research activities and attract more attention to VaaS in both academia and industries to develop an alternative economically sound approach to building smart cities. To push this paradigm into reality, we urgently call for the joint coordinated and concerted efforts of all governmental branches, communication and computing industries, AI sectors, and research communities.

\renewcommand\refname{References}
\bibliographystyle{IEEEtran}
\bibliography{IEEEabrv,fang,reference}

\vspace{-33pt}
\begin{IEEEbiography}[{\includegraphics[width=1in,height=1.25in,clip,keepaspectratio]{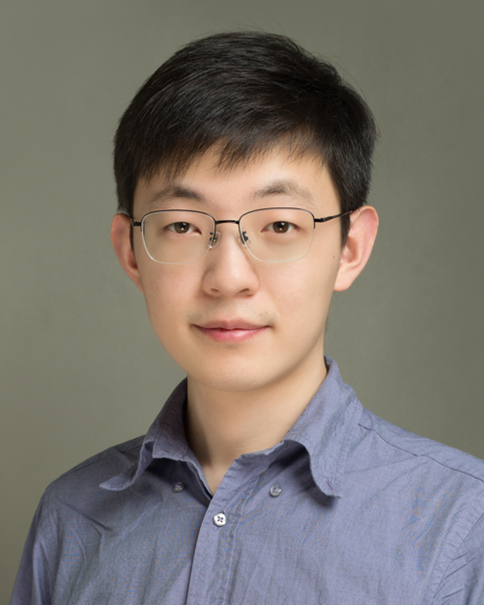}}]{Xianhao Chen}(Member, IEEE) received the B.Eng. degree in electronic information from Southwest Jiaotong University in 2017, and the Ph.D. degree in electrical and computer engineering from the University of Florida in 2022. He is currently an assistant professor with the Department of Electrical and Electronic Engineering, the University of Hong Kong. He received the 2022 ECE graduate excellence award for research from the University of Florida. His research interests include wireless networking, edge intelligence, and machine learning.
\end{IEEEbiography}

\vspace{-33pt}
\begin{IEEEbiography}[{\includegraphics[width=1in,height=1.25in,clip,keepaspectratio]{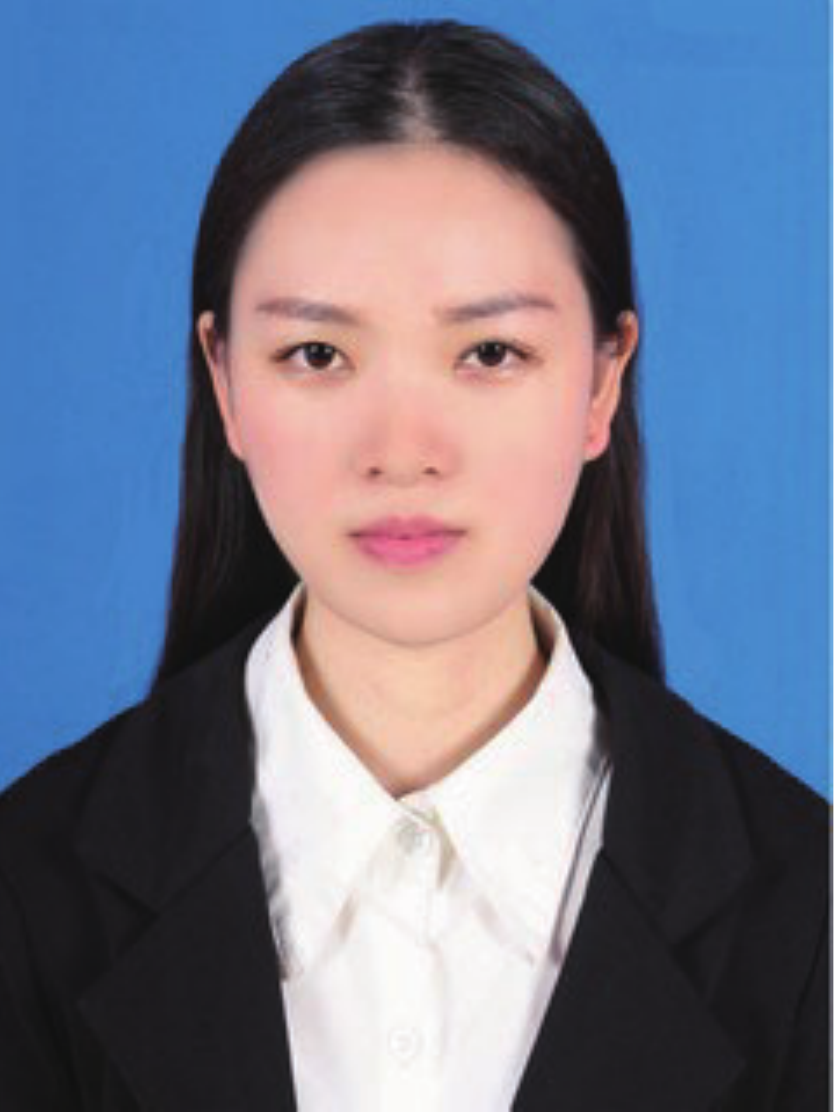}}]{Yiqin Deng}(Member, IEEE) received her B.S. degree in project management from Hunan Institute of Engineering, Xiangtan, China, in 2014, and her M.S. degree in software engineering and her Ph.D. degree in computer science and technology from Central South University, Changsha, China, in 2017 and 2022, respectively. She is currently a Postdoctoral Research Fellow with the School of Control Science and Engineering, Shandong University, Jinan, China. She was a visiting researcher at the University of Florida, Gainesville, from 2019 to 2021. Her research interests include edge/fog computing, Internet of Vehicles, and resource management.
\end{IEEEbiography}

\vspace{-33pt}
\begin{IEEEbiography}[{\includegraphics[width=1in,height=1.25in,clip,keepaspectratio]{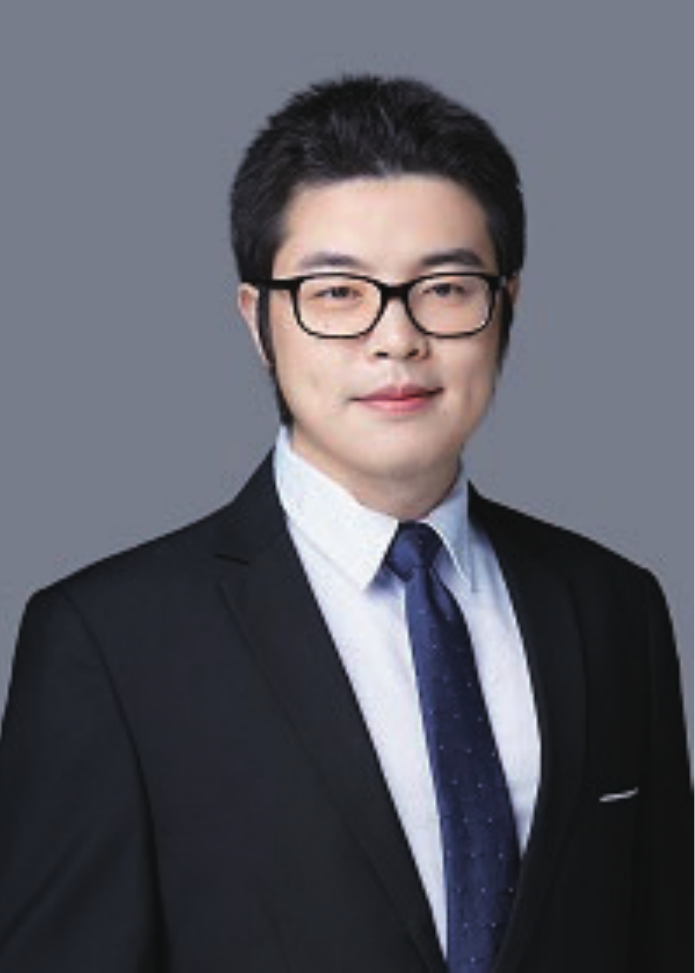}}]{Haichuan Ding}(Member, IEEE)
received the B.Eng. and M.S. degrees in electrical engineering from the Beijing Institute of Technology (BIT), Beijing, China, in 2011 and 2014, respectively, and the Ph.D. degree in electrical and computer engineering from the University of Florida, Gainesville, FL, USA, in 2018. He is currently a professor with the School of Cyberspace Science and Technology, Beijing Institute of Technology, Beijing, China. His research interests include cognitive radio networks, edge communications/computing, covert communications and space-air-ground integrated networks.
\end{IEEEbiography}

\vspace{-33pt}
\begin{IEEEbiography}[{\includegraphics[width=1in,height=1.25in,clip,keepaspectratio]{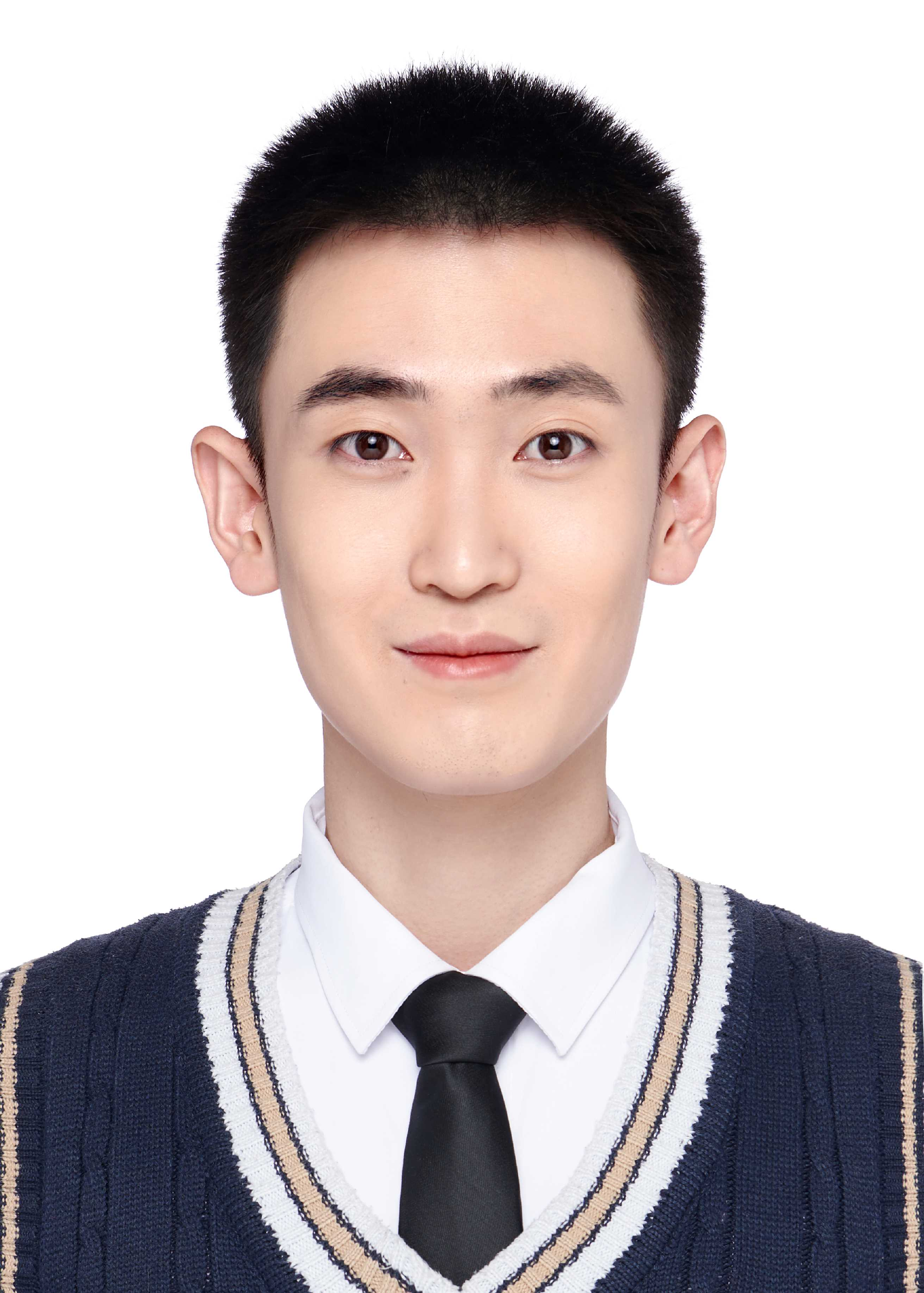}}]{Guanqiao Qu}(Graduate Student Member, IEEE)
received the B.E. and M.E. degrees from the Harbin Institute of Technology (HIT), Harbin, China, in 2020 and 2022, respectively. He is currently pursuing the Ph.D. degree with the Department of Electrical and Electronic Engineering, the University of Hong Kong (HKU), Hong Kong, China. His research interests include wireless communications, networking, edge computing, vehicular ad hoc networks, federated learning, and resource allocation. 
\end{IEEEbiography}

\vspace{-33pt}
\begin{IEEEbiography}
[{\includegraphics[width=1 in, height=1.25 in, clip, keepaspectratio]{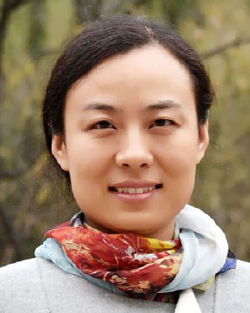}}] 
{Haixia Zhang} (Senior Member, IEEE) received the B.E. degree from the Department of Communication and Information Engineering, Guilin University of Electronic Technology, Guilin, China, in 2001, and the M.Eng. and Ph.D. degrees in communication and information systems from the School of Information Science and Engineering, Shandong University, Jinan, China, in 2004 and 2008, respectively. 

From 2006 to 2008, she was with the Institute for Circuit and Signal Processing, Munich University of Technology, Munich, Germany, as an Academic Assistant. From 2016 to 2017, she was a Visiting Professor with the University of Florida, Gainesville, FL, USA. She is currently a Full Professor with Shandong University, Jinan, China. Dr. Zhang is actively participating in many professional services. She is an editor of the IEEE Transactions on Wireless Communications, IEEE Wireless Communication Letters, and China Communications and serves/served as  Symposium Chairs, TPC Members, Session Chairs, and Keynote Speakers of many conferences. Her research interests include wireless communication and networks, industrial Internet of Things, wireless resource management, and mobile edge computing. 
\end{IEEEbiography}

\vspace{-33pt}
\begin{IEEEbiography}
[{\includegraphics[width=1 in, height=1.25 in, clip, keepaspectratio]{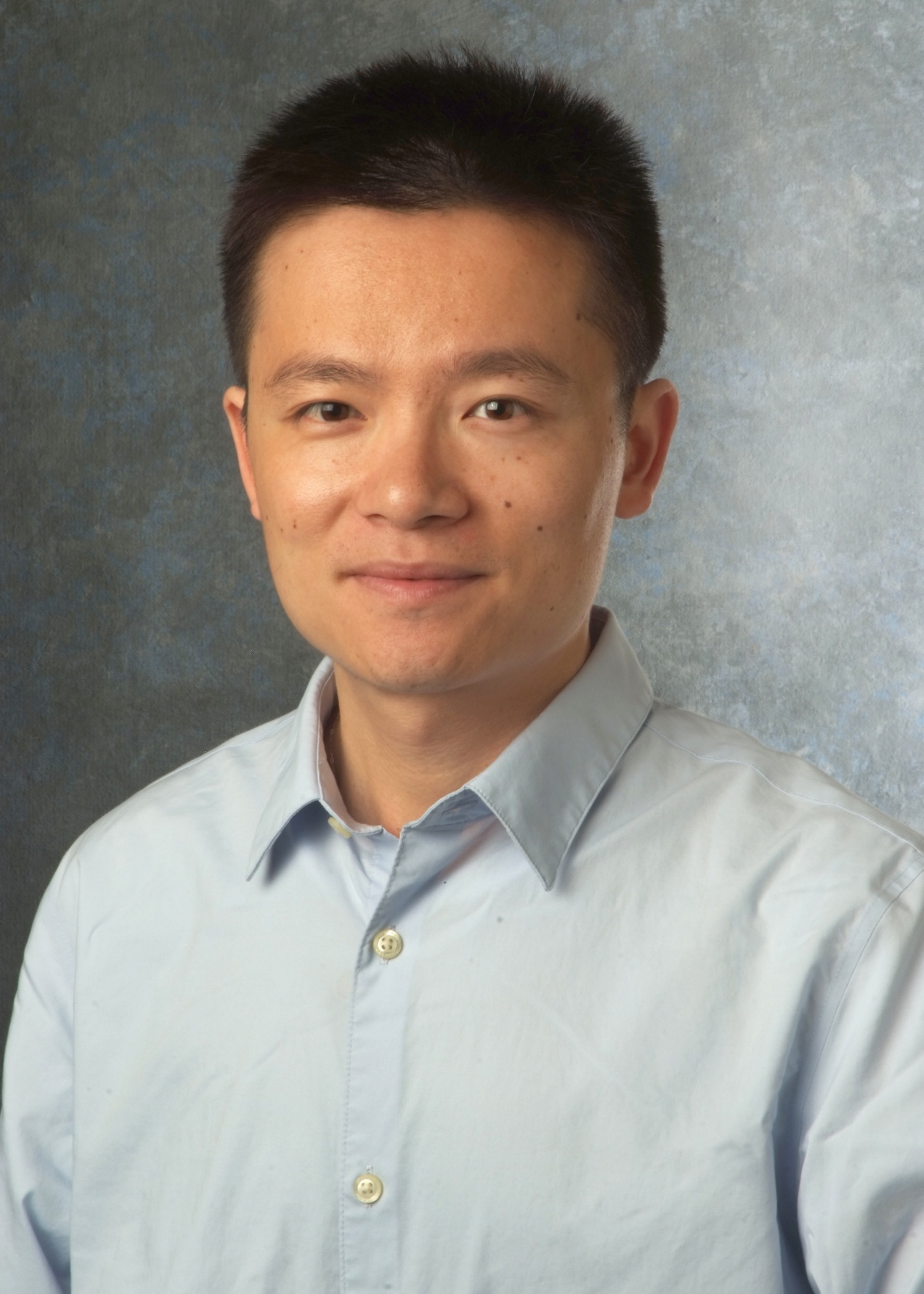}}] 
{Pan Li} (Senior Member, IEEE) received the B.E. degree in electrical engineering from Huazhong University of Science and Technology, Wuhan, China, in 2005, and the Ph.D. degree in electrical and computer engineering from University of Florida, Gainesville, Florida, USA in 2009, respectively. He was an Assistant Professor in the Department of Electrical and Computer Engineering at Mississippi State University from 2009 to 2015. Since Fall 2015, he has been with the Department of Electrical Engineering and Computer Science at Case Western Reserve University, Cleveland, Ohio, USA. 

Dr. Li received the US NSF CAREER Award in 2012. He has been serving/served as an Editor of IEEE Transactions on Big Data, IEEE Transactions on Mobile Computing, IEEE Transactions on Network Science and Engineering, IEEE Transactions on Wireless Communications, IEEE Wireless Communications Letters, IEEE Journal on Selected Areas in Communications–Cognitive Radio Series, IEEE Communications Surveys and Tutorials, Computer Networks, and a Feature Editor of IEEE Wireless Communications. He has also served on the organizing committee and technical program committee of flagship conferences such as IEEE ICC, IEEE INFOCOM, ACM MobiHoc, AAAI, IJCAI. His research interests include machine learning, cybersecurity, and network science. 
\end{IEEEbiography}

\vspace{-33pt}
\begin{IEEEbiography}[{\includegraphics[width=1in,height=1.25in,clip,keepaspectratio]{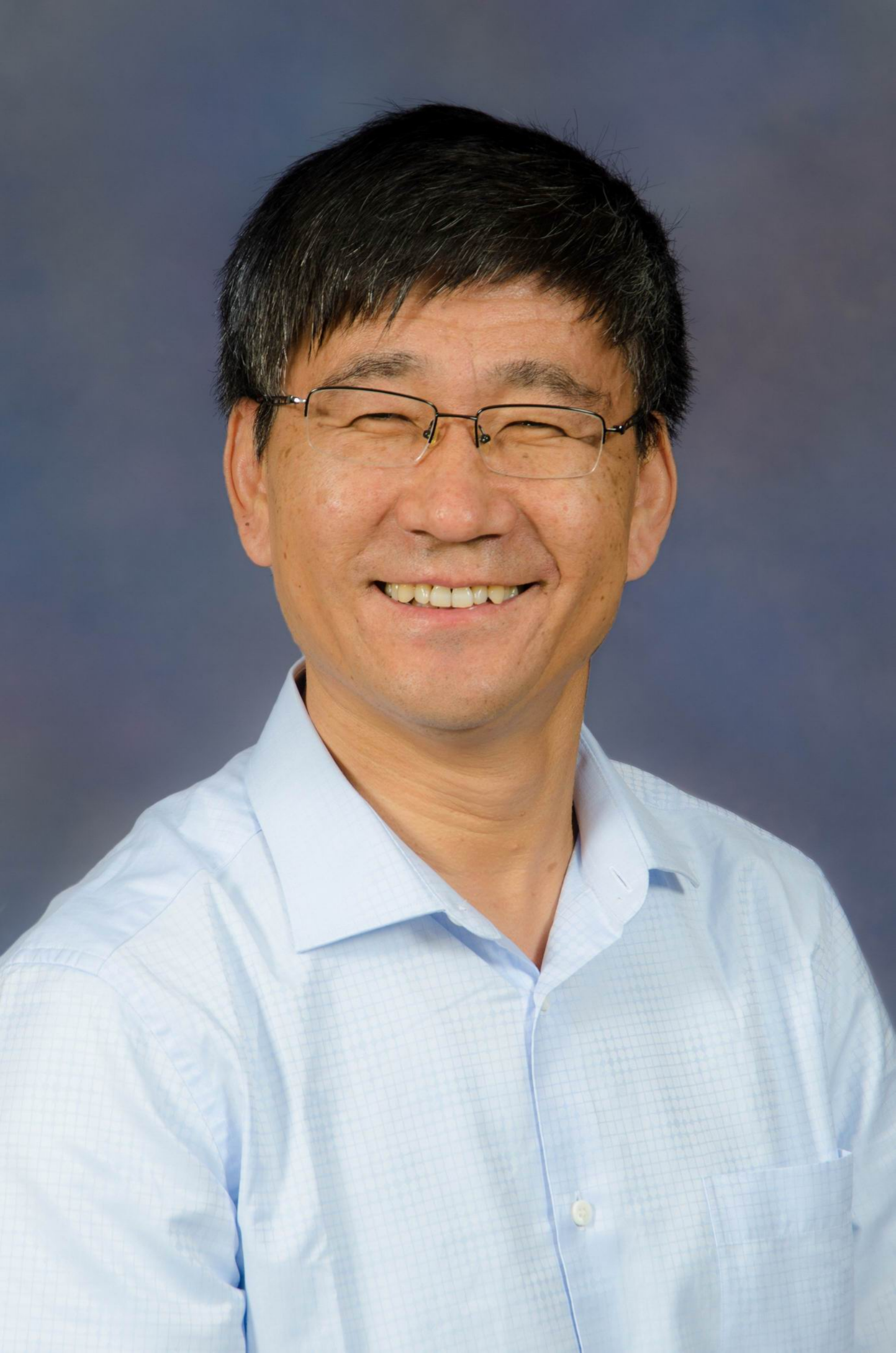}}]{Yuguang Fang}(Fellow, IEEE)
received the MS degree from Qufu Normal University, China in 1987, a PhD degree from Case Western Reserve University, Cleveland, Ohio, USA, in 1994, and a PhD degree from Boston University, Boston, Massachusetts, USA in 1997. He joined the Department of Electrical and Computer Engineering at University of Florida in 2000 as an assistant professor, then was promoted to associate professor in 2003, full professor in 2005, and distinguished professor in 2019, respectively. Since August 2022, he has been the Chair Professor of Internet of Things with the Department of Computer Science at City University of Hong Kong.

Dr. Fang received many awards including the US NSF CAREER Award (2001), US ONR Young Investigator Award (2002), 2018 IEEE Vehicular Technology Outstanding Service Award, 2019 IEEE Communications Society AHSN Technical Achievement Award, 2015 IEEE Communications Society CISTC Technical Recognition Award, 2014 IEEE Communications Society WTC Recognition Award, the Best Paper Award from IEEE ICNP (2006), 2010-2011 UF Doctoral Dissertation Advisor/Mentoring Award, and 2009 UF College of Engineering Faculty Mentoring Award. He held multiple professorships including the Changjiang Scholar Chair Professorship (2008-2011), Tsinghua University Guest Chair Professorship (2009-2012), NSC Visiting Researcher of National Taiwan University (2007-2008), Invitational Fellowship of Japan Society for the Promotion of Science (2009), University of Florida Foundation Preeminence Term Professorship (2019-2022), University of Florida Research Foundation Professorship (2017-2020, 2006-2009), and University of Florida Term Professorship (2017-2021). He served as the Editor-in-Chief of IEEE Transactions on Vehicular Technology (2013-2017) and IEEE Wireless Communications (2009-2012) and serves/served on several editorial boards of journals including Proceedings of the IEEE (2018-present), ACM Computing Surveys (2017-present), ACM Transactions on Cyber-Physical Systems (2020-present), IEEE Transactions on Mobile Computing (2003-2008, 2011-2016, 2019-present), IEEE Transactions on Communications (2000-2011), and IEEE Transactions on Wireless Communications (2002-2009). He served as the Technical Program Co-Chair of IEEE INFOCOM’2014 and the Technical Program Vice-Chair of IEEE INFOCOM'2005. He has actively engaged with his professional community, serving as a Member-at-Large of the Board of Governors of IEEE Communications Society (2022-2024) and the Director of Magazines of IEEE Communications Society (2018-2019). He is a fellow of ACM, IEEE, and AAAS.
\end{IEEEbiography}

\end{document}